\documentclass[preprint,authoryear]{elsarticle}

\usepackage[utf8]{inputenc}             
\usepackage{amsmath, amssymb}           
\usepackage{graphicx}                   
\usepackage{pdflscape}
\usepackage{array}
\usepackage{float}                      
\usepackage{longtable}                  
\usepackage{colortbl, xcolor}           
\usepackage{tabularx}                   
\usepackage{ragged2e}                   
\usepackage{svg}                        
\usepackage{adjustbox}                  
\usepackage{multirow}                   
\usepackage{comment}                    
\usepackage{soul}                       
\usepackage[a4paper, margin=0.75in]{geometry} 
\usepackage[linesnumbered,ruled,vlined]{algorithm2e} 
\usepackage{booktabs}
\usepackage{chngcntr}
\counterwithout{table}{section}

\rowcolors{2}{gray!5}{white}
\usepackage[authoryear]{natbib}         
\usepackage{tcolorbox}

\journal{Transportation Research Interdisclipinary Perspectives}

\begin{document}

\begin{frontmatter}

\title{Reasons and Principles for Automated Vehicle Decisions in Ethically Ambiguous Everyday Scenarios}

\author[inst1,inst3]{Lucas Elbert Suryana}
\author[inst1,inst3]{Simeon Calvert}
\author[inst3,inst2]{Arkady Zgonnikov}
\author[inst1]{Bart van Arem}

\affiliation[inst1]{organization={Department of Transport and Planning, Faculty of Civil Engineering and Geosciences, Delft University of Technology},
            city={Delft},
            country={The Netherlands}}

\affiliation[inst2]{organization={Department of Cognitive Robotics, Faculty of Mechanical Engineering, Delft University of Technology},
            city={Delft},
            country={The Netherlands}}

\affiliation[inst3]{organization={Centre for Meaningful Human Control, Delft University of Technology},
            city={Delft},
            country={The Netherlands}}

\begin{abstract}
Automated vehicles (AVs) increasingly encounter ethically ambiguous situations in everyday driving—scenarios involving conflicting human interests and lacking clearly optimal courses of action. While existing ethical models often focus on rare, high-stakes dilemmas (e.g., crash avoidance or trolley problems), routine decisions such as overtaking cyclists or navigating social interactions remain underexplored. This study addresses that gap by applying the tracking condition of Meaningful Human Control (MHC), which holds that AV behaviour should align with human reasons—defined as the values, intentions, and expectations that justify actions. We conducted qualitative interviews with 18 AV experts to identify the types of reasons that should inform AV manoeuvre planning. Thirteen categories of reasons emerged, organised across normative, strategic, tactical, and operational levels, and linked to the roles of relevant human agents. A case study on cyclist overtaking illustrates how these reasons interact in context, revealing a consistent prioritisation of safety, contextual flexibility regarding regulatory compliance, and nuanced trade-offs involving efficiency, comfort, and public acceptance. Based on these insights, we propose a principled conceptual framework for AV decision-making in routine, ethically ambiguous scenarios. The framework supports dynamic, human-aligned behaviour by prioritising safety, allowing pragmatic actions when strict legal adherence would undermine key values, and enabling constrained deviations when appropriately justified. This empirically grounded approach advances current guidance by offering actionable, context-sensitive design principles for ethically aligned AV systems.
\end{abstract}


\end{frontmatter}


\newpage

\section{Introduction}
\label{sec:introduction}

As automated vehicle (AV) technology continues to advance, significant challenges remain, particularly when AVs must make decisions in ethically complex situations \citep{nyholm2016ethics, wang2020ethical, Saber2024}. These situations arise when AVs are required to balance safety, efficiency, and compliance with societal expectations—ranging from minimizing risks for all road users to navigating dilemmas involving conflicting values, interests, or trade-offs. Such conflicts often create ambiguity about the most appropriate course of action for AVs \citep{himmelreich2018never, bergmann2022ethical}. Addressing these dilemmas requires not only technical advancements, especially in planning and decision-making \citep{schwarting2018planning, geisslinger2023ethical}, but also the development of clearer ethical principles.

\subsection{Ethically straightforward guidelines}
\label{Ethically_straightforward_guidelines}
Stakeholders involved in AV development and regulation have proposed various recommendations and guidelines for how AVs should behave in safety-critical situations. Among these, the concept of roadmanship has been introduced as a guiding principle to ensure that AVs drive safely, avoid creating hazards, and respond effectively to hazards caused by others \citep{fraadeblanar2018measuring}. Although less formalised than regulatory guidelines, roadmanship emphasizes predictability and anticipatability in driving behaviour, similar to how a competent and careful human driver navigates traffic.

This concept aligns with the UNECE guidelines \citep{UNECE2023}, which present reference models of “competent and careful” drivers. These models serve as benchmarks for evaluating AV behaviour in safety-critical scenarios such as cut-ins, cut-outs, and lead-vehicle deceleration, reflecting how a skilled human driver would minimise risk. If an AV outperforms these reference models, it is considered safer than a competent and careful human driver. Performance metrics include time headway, time-to-collision, and vehicle positioning, regardless of whether an accident is preventable \citep{olleja2025validation}.

However, the safety-critical situations addressed by these models are ethically straightforward, since all parties benefit from the AV’s behaviour. For example, the UNECE scenarios focus on mutually beneficial outcomes, such as avoiding collisions and reducing risk for all traffic participants. While these benchmarks offer important guidance for critical events, they do not address the full spectrum of ethically challenging situations that AVs may encounter in everyday driving.

\subsection{Ethically ambiguous guidelines}
\label{Ethically_ambiguous_guidelines}
In everyday driving, AVs frequently encounter ambiguous situations involving conflicting interests, where the optimal course of action is unclear. Routine traffic scenarios—such as approaching a crosswalk with limited visibility or making a left turn in the presence of oncoming traffic—exemplify such dilemmas \citep{himmelreich2018never}. These situations often require balancing safety, legal compliance, and traffic efficiency. For instance, when approaching a crosswalk with limited visibility, an AV must decide whether to prioritise slowing down to ensure pedestrian safety, at the expense of reducing traffic flow, or maintain speed to optimise mobility, potentially increasing risk.

Unlike human drivers, who make such decisions intuitively on a case-by-case basis, drawing on experience and an understanding of human behaviour, AVs must encode these trade-offs systematically across all vehicles. This raises a fundamental ethical challenge: how should AVs navigate scenarios where competing priorities conflict at a systemic level?

Some researchers have examined these challenges through the lens of the trolley problem, which explores moral judgements in life-and-death trade-offs. One approach is rooted in consequentialist ethics, where actions are evaluated based on outcomes. For example, \citet{meder2019should} uncovered nuanced ways in which individuals’ moral judgements reflect consequentialist reasoning. In contrast, deontological perspectives prioritise adherence to moral principles, such as the protection of passengers. \citet{liu2021selfish} found that participants favoured AVs programmed to protect their occupants at all costs. Complementing these, the MIT Moral Machine project adopted a virtue-based approach, highlighting significant cultural variation in ethical preferences for AV decision-making \citep{awad2018moral}.

While these studies reveal diverse ethical perspectives, they also underscore the difficulty of developing a universal framework for AV behaviour. Critics argue, however, that the trolley problem has limited relevance to real-world driving, as AVs are unlikely to encounter such stark life-and-death scenarios during routine operation \citep{nyholm2016ethics, keeling2020trolley}. Instead, AVs more commonly face ethically ambiguous situations where the stakes are lower but the complexity is higher—making these scenarios critical for AV development and deployment \citep{himmelreich2018never}.

Recent recommendations from a European Commission Expert Group propose guidelines for addressing crash dilemmas through risk distribution and shared ethical principles \citep{bonnefon2020ethics}. Although these recommendations are valuable for ethically complex crash scenarios, they assume that crashes are unavoidable and do not fully address the challenges posed by routine, ethically ambiguous situations. This highlights the need for a more structured approach to navigating everyday ethical challenges, emphasising the importance of identifying principles that should guide AV behaviour in such contexts.

\subsection{Guidelines based on the concept of Meaningful Human Control}
\label{Guidelines_MHC}

The concept of Meaningful Human Control (MHC) emerged in response to concerns over wrongful actions by automated weapon systems, which could create a responsibility gap \citep{asaro2012banning, horowitz2015}. Without meaningful human control, such systems might make life-and-death decisions that result in unintended casualties or violations of international law, such as targeting civilians rather than enemy combatants. Similarly, in the context of AVs, the failure to resolve ethical dilemmas—such as balancing pedestrian safety with traffic flow—could lead to safety-critical failures.

\citet{santoni2018meaningful} subsequently developed a foundational theory defining what MHC entails. According to this theory, automated systems should be responsive to the reasons provided by human agents for their decisions, ensuring alignment with human values and intentions—a principle referred to as the tracking condition of MHC. Researchers have since proposed ways to operationalise MHC for AVs. For instance, \citet{mecacci2020meaningful} explores how the concept can be applied to the AV domain, while \citet{calvert2020conceptual} builds a conceptual model for implementing MHC in the control systems of connected and autonomous vehicles (CAVs) in mixed traffic environments.

We argue that the tracking condition of MHC provides a suitable foundation for developing guidelines to support AV behaviour in ethically ambiguous routine driving scenarios. Unlike rigid ethical frameworks based on predefined rules (e.g., utilitarian or consequentialist principles), MHC emphasises understanding and incorporating the reasoning behind human decisions. In such scenarios, human drivers often know how to act, relying on their ability to interpret context, weigh competing considerations, and make judgements accordingly. This natural adaptability underscores the importance of designing AVs that can dynamically respond to human reasoning in specific contexts. By focusing on the tracking of human reasons, MHC enables AVs to align their actions with human values and intent—an essential capability in real-world driving environments where ambiguity and unpredictability are common.

\subsection{Research Gaps and Objectives}

\begin{itemize}
\item \textbf{Theoretical gap}: While several AV ethics frameworks—such as the Integrative Ethical Decision-Making Framework \citep{rhim2021deeper}—have been proposed to address moral dilemmas, they primarily focus on high-stakes or exceptional crash scenarios. As a result, they often overlook the ethical complexity inherent in routine driving situations. The concept of MHC, particularly its tracking condition \citep{santoni2018meaningful}, offers a promising basis for addressing this gap by facilitating alignment between AV behaviour and human reasoning in everyday contexts. However, its application remains underexplored in ethically ambiguous routine scenarios. This study contributes by operationalising MHC to inform AV decision-making in such contexts.
\item \textbf{Practical gap}: Although there is growing recognition that AVs must exhibit socially sensitive behaviour— that is, behaviour aligned with human values, expectations, and contextually appropriate responses \citep{d2022exceptional, lu2025empowering}—current frameworks lack a structured, empirically grounded set of human reasons that AVs should consider during ethically complex, routine situations (e.g., overtaking a cyclist). This study addresses this gap by developing a principled framework that categorises expert-elicited reasons across normative, strategic, tactical, and operational levels of AV behaviour.

\item \textbf{Methodological gap}: Existing research often relies on simulation models or moral vignettes, with limited empirical input from domain experts \citep{dubljevic2023moral}. While some studies incorporate expert perspectives in extreme scenarios \citep{milford2025all}, few systematically capture expert reasoning relevant to the daily ethically ambiguous conditions AVs encounter in everyday real-world operation. This study fills that gap through qualitative interviews with 18 AV experts, using a structured analysis informed by the MHC framework.
\end{itemize}

Building on the tracking condition of MHC, this study aims to address these gaps by proposing a methodology that can inform and complement existing guidelines. To achieve this, the study has two primary objectives:
\begin{itemize}
    \item \textbf{Objective 1}: To gather the reasons provided by AV experts regarding the factors they believe AVs should consider when planning a manoeuvre.
    \item \textbf{Objective 2}: To develop a principled conceptual framework for AV behaviour in ethically challenging routine situations, based on the reasons identified in Objective 1.
\end{itemize}

To address Objective 2, we adopt a two-step approach. In the first step, we classify the reasons elicited from AV experts into groups, aiming to identify underlying principles. In the second step, a case study of an ethically ambiguous routine scenario—such as overtaking a cyclist—is presented to the experts. They are asked to provide recommendations on the manoeuvre decisions AVs should make in such situations and to explain the reasoning behind their recommendations. These reasons are then mapped back to the classifications from the first step to uncover relationships and derive expert-informed guidelines for AV behaviour.

\section{Expert Participant Description}

\subsection{Selection Criteria}
To ensure that the study reflected insights from individuals with substantive knowledge of AV systems, we employed a purposive sampling strategy, complemented by snowball sampling. Initial participants were selected through the professional networks of the authors and evaluated based on their publication records, institutional affiliations, and topic relevance, as reflected in publicly available sources such as Google Scholar profiles. This approach enabled the identification of experts with demonstrable contributions to AV-related research and development, in line with accepted practices in qualitative transportation studies. For example, \citet{ma2024analysing} recruited AV professionals through LinkedIn based on their hands-on experience with automated systems, while \citet{hilgarter2020public} employed purposive sampling in a real-world AV deployment by selecting participants immediately after they experienced an autonomous shuttle ride—ensuring high contextual relevance. Similarly, our approach aimed to ensure that participants had domain-specific expertise in AVs and were capable of contributing informed reasoning about AV decision-making.

Subsequent participants were recruited via expert referrals following early interviews. This snowball sampling method enabled us to reach additional individuals working in specialised domains who may not have been immediately visible through conventional directories. Comparable combined strategies have been used in AV-focused qualitative studies to capture diverse, high-level perspectives from academia, industry, and government \citep{milford2025all}.

\subsection{Recruitment}
Participants were recruited via personalised email invitations. Each email included a brief overview of the study, highlighting its focus on understanding trade-offs in motion planning for overtaking scenarios involving automated vehicles (AVs). We clarified that although the study focused on motion planning, participation was not limited to specialists in that area; instead, we sought a broad range of perspectives from individuals involved in AV ethics, design, policy, and engineering.

The email also outlined the interview format and logistics: semi-structured, approximately 45–60 minutes in duration, conducted via Zoom, and optionally recorded with participant consent. The voluntary nature of participation and the right to withdraw at any time were clearly communicated. Recruitment and study procedures were approved by the Human Research Ethics Committee (HREC) of Delft University of Technology (ID: 132530).

\subsection{Participant Profile}
The final sample consisted of 18 expert participants from seven countries, representing a range of perspectives on AV development. Of the 35 experts initially contacted—14 from the United States, 13 from the Netherlands, four from the United Kingdom, and one each from Italy, Belgium, Israel, and Japan—18 agreed to participate, resulting in a response rate of 51

The participant profile reflects diverse institutional affiliations and technical backgrounds. As shown in Table \ref{tab:expertise_positions}, participants were drawn from academia (n = 12) and industry (n = 6), with disciplinary expertise in motion planning, human factors, ethics, behavioural science, and legal policy. Based on self-reported experience, participants had on average more than five years of direct involvement with automated vehicle development. This diversity of expertise and roles contributed to a rich and multidimensional set of perspectives on AV decision-making in motion planning contexts.

\subsection{Procedure}
Participants were interviewed online using Microsoft Teams, with audio, video, and transcription recorded. The interviews followed a predefined protocol consisting of both open-ended and closed-ended questions. To minimise potential interviewer bias and ensure consistency, the questionnaire was developed using Qualtrics (http://www.qualtrics.com), and participants received a direct link via chat at the beginning of the session. This enabled them to independently read and progress through the questions at their own pace.

During the interviews, the researcher primarily observed and listened, intervening minimally and only providing clarifications when explicitly requested by participants. Participants had the flexibility to skip questions if they felt these had already been adequately addressed. This procedure aligns with best practices for semi-structured qualitative interviewing in AV research, where researcher involvement is intentionally limited to preserve objectivity and ensure consistency across interviews \citep{nordhoff2023mis}.

\begin{table}[H]
\small
\centering
\caption{Overview of Expert Participants by Sector, Country, and Expertise}
\label{tab:expertise_positions}
\begin{adjustbox}{max width=\textwidth}
\begin{tabular}{llp{6cm}l}
\toprule
\textbf{ID} & \textbf{Country} & \textbf{Expertise} & \textbf{Role} \\
\midrule
\multicolumn{4}{l}{\textit{Academia}} \\
1 & Netherlands & Human-AI interaction, ethics & Researcher \\
2 & US          & Technical validation, travel behaviour & Researcher \\
3 & US          & AV safety validation & Researcher \\
4 & Netherlands & Motion planning algorithms & Researcher \\
5 & Netherlands & Road users and infrastructure perspectives & Researcher \\
6 & Netherlands & Ethics of AI & Researcher \\
7 & UK          & Modelling human behaviour & Researcher \\
8 & Netherlands & Social science of behaviour & Researcher \\
9 & UK          & AV user experience & Researcher \\
10 & Israel     & Public perception and AV ethics & Researcher \\
11 & UK         & Human factors in transport & Researcher \\
12 & Netherlands & Legal aspects of AV & Researcher \\
\midrule
\multicolumn{4}{l}{\textit{Industry}} \\
13 & UK         & AV safety and assurance & Consultant \\
14 & Netherlands & Traffic psychology & Psychologist \\
15 & US         & Software quality assurance & Engineer \\
16 & US         & Driving strategy, business development & Consultant \\
17 & US         & AV safety & Consultant \\
18 & US         & Human factors & Researcher \\
\bottomrule
\end{tabular}
\end{adjustbox}
\end{table}

The interview protocol consisted of four main parts:
\begin{itemize}
    \item \textbf{Part 1 (Q2–Q4)}: Open-ended questions exploring factors that influence AV manoeuvre planning.
    \item \textbf{Part 2 (Q5–Q9)}: Evaluation of an ethically challenging real-world AV scenario involving a cyclist, asking participants to identify and assess reasons relevant to decision-making.
    \item \textbf{Part 3 (Q10–Q13)}: Ranking of predefined reasons, enabling participants to indicate the most appropriate decision in the given scenario.
    \item \textbf{Part 4 (Q14–Q19)}: Evaluation of alternative AV decisions, using time-based assessments to examine how stakeholder reasons were addressed in a revised scenario.
\end{itemize}

Finally, participants discussed abstract considerations regarding how AVs might interpret stakeholder intentions and manage potential conflicts (Q20–Q21).

This paper analyses responses from Q2–Q4 (addressing Objective 1) and Q5–Q9 (in conjunction with outcomes from Objective 1, addressing Objective 2), as outlined in Table~\ref{tab:interview_questions}. Table A1 in the Appendix presents the full set of questions used during the interviews. Analysis of the remaining questions lies beyond the scope of this study and will be addressed in future work.

\begin{table}[ht]
\centering
\caption{Interview Questions Relevant to Objectives 1 and 2}
\begin{tabular}{|l|p{13.5cm}|}
\hline
\rowcolor[gray]{0.9}
\textbf{Question Number} & \textbf{Question} \\
\hline
Q2 & What should automated vehicles (AVs) consider when planning a maneuver? Please give one example in as much detail as possible. \\
\hline
Q3 & Which moral aspects do you believe AVs should consider when planning a maneuver? \\
\hline
Q4 & How might these aspects affect the maneuver plan? \\
\hline
\multicolumn{2}{|p{\dimexpr\linewidth-2\tabcolsep\relax}|}{
\textbf{Please watch the video below and read its description} \newline
\textbf{Video clip (see ~\ref{sec:appendix:video})} \newline
\textbf{Video description}
\newline
A passenger uses an automated vehicle (AV) for a morning commute to the office. The passenger has an important meeting and must arrive on time. If the vehicle maintains the current speed, the passenger can reach the office on time in 20 minutes. The AV is on a road with solid double yellow lines, which prohibit vehicles from crossing in both directions due to safety reasons. During the trip, the AV approaches a cyclist traveling at half of the speed of the AV. There is no safe passing zone visible from the vehicle; however, the opposite lane is currently empty.  
} \\
\hline
Q5 & If the video continues, what do you believe all traffic participants will do next? \\
\hline
Q6 & What are the reasons for the traffic participants performing the actions you mentioned? \\
\hline
Q7 & Besides those traffic participants, can you think of any other factors that might influence their decisions? \\
\hline
Q8 & What do you think the reasons are for the other factors you mentioned (e.g., traffic signs and the double yellow line)? \\
\hline
Q9 & Can you think of any situations where the intentions of the [traffic participants / other factors the experts mentioned\ might conflict? Please share any examples you can think of, and let me know when these conflicts may typically occur.] \\
\hline
\end{tabular}
\label{tab:interview_questions}
\end{table}

\section{Reasons That Influence Considerations for AV Manoeuvre Planning}
\label{sec:principled_framework}

\subsection{Questionnaire Design}
\label{questionnaire_design}

The questionnaire used in this research was designed as a semi-structured instrument, informed by the tracking condition of the Meaningful Human Control (MHC) framework. This condition holds that automated systems should respond to relevant human reasons \citep{santoni2018meaningful}. In this article, we define “reasons” as normative reasons or factual considerations that justify particular actions, rather than motivational reasons, following the distinction outlined by \citet{veluwenkamp2022reasons}.

According to the MHC framework, reasons relevant to automated vehicle (AV) decision-making can be grouped into four categories: moral, strategic, tactical, and operational. Moral reasons pertain to ethical principles or social norms (e.g., fairness, harm avoidance). Strategic reasons relate to long-term planning goals (e.g., minimising travel time). Tactical reasons involve interactions with other road users (e.g., overtaking or yielding), while operational reasons concern real-time control actions (e.g., braking, steering).

To elicit expert views on these types of reasons, participants were asked a series of open-ended questions designed to explore what considerations AVs should take into account when planning a manoeuvre. Specifically, Questions 2–4 (see Table~\ref{tab:interview_questions}) focused on eliciting examples and explanations of factors that experts deemed important in AV planning contexts. These responses formed the basis for a theory-driven qualitative coding process aimed at identifying the types of reasons referenced and mapping them to the four categories outlined in the MHC framework. The analysis procedure is detailed in Section~3.2.

\subsection{Data Analysis}
\subsubsection{Coding Framework}

We used directed content analysis to analyse the expert responses, applying the Meaningful Human Control (MHC) framework as a coding guide \citep{mecacci2020meaningful}. The framework classifies reasons into four types based on their level of abstraction, timing, complexity, and the role of the decision-maker:

\begin{itemize}
    \item \textbf{Moral reasons:} Ethical principles or social/legal norms (e.g., fairness, traffic law compliance). These are abstract, long-term in scope, and typically shaped by institutions or broader societal expectations.
    \item \textbf{Strategic reasons:} Broader planning goals such as route optimisation or travel efficiency. These are moderately abstract, span longer durations, and are usually attributed to the driver as planner.
    \item \textbf{Tactical reasons:} Short-term decisions made during driving, such as merging or overtaking. These are more concrete and informed by the immediate driving context.
    \item \textbf{Operational reasons:} Real-time control actions, including steering and braking. These are highly specific and implemented by the AV system or human driver in response to moment-to-moment environmental cues.
\end{itemize}

We created a coding matrix based on these categories. Each response to Questions 2–4 was segmented into individual statements, which were then coded according to the type of reason expressed. Statements could be assigned to one or more MHC categories, based on their content, timing, and the type of agent typically responsible for such decisions.

This process enabled us to identify not only which categories of reasons were most frequently cited by experts, but also how these reasons varied across different driving contexts. For example, a participant might refer to traffic law compliance (moral), route efficiency (strategic), minimising lane changes (tactical), or maintaining a safe following distance (operational).

When statements reflected more than one type of reason, we used cross-coding to preserve their complexity and ensure comprehensive coverage. This method allowed us to retain the nuance of expert input while maintaining a structured ethical and functional classification scheme.

Our approach, grounded in the theoretical framework of MHC, is consistent with other AV studies employing theory-driven content analysis. For instance, \citet{aasvik2025trust} used a similar method to analyse public trust in autonomous shuttles, while \citet{suryana2025} applied the MHC framework to explore interview data in relation to the tracking and tracing conditions.

\subsubsection{Qualitative Content Analysis Procedure}
To apply the Meaningful Human Control (MHC) framework in a structured and transparent manner, we conducted a qualitative content analysis that combined theory-driven and data-driven steps. We began by segmenting interview transcripts into individual response units. Each unit was analysed to identify four key components: (1) the AV behaviour being recommended (consideration), (2) the justification for that behaviour (reason), (3) the human agent associated with the reason, and (4) the corresponding MHC category—moral, strategic, tactical, or operational.

We explicitly distinguished between considerations and reasons. Considerations refer to the specific behaviour that the AV is expected to perform (e.g., “the AV should slow down near pedestrians”), whereas reasons are the human-oriented justifications for those behaviours (e.g., “to ensure the safety of vulnerable road users”). Each reason was then evaluated according to the MHC criteria. This involved examining its temporal relevance (whether it referred to planning, real-time action, or reflection), its level of abstraction (concrete vs general), and the role of the human agent to whom the reason was attributed (e.g., designer, operator, policymaker). When a reason encompassed multiple types of justification—such as combining moral fairness with strategic efficiency—it was assigned to more than one MHC category.

Our approach recognised that reasons could be either explicitly stated in the data or logically inferred from context. Explicit reasons were identified when participants directly articulated the justification for their statements. In other cases, implicit reasons were inferred based on the surrounding narrative. This approach draws on principles of latent content analysis, in which underlying meanings are interpreted beyond the literal language used. Latent content, as defined by \citet{graneheim2004qualitative}, refers to the deeper meaning embedded in a text—especially important when participants allude to motivations or norms without stating them directly. Building on this, scholars such as \citet{vaismoradi2013content} and \citet{krippendorff2018content} have emphasised how interpreting latent content can uncover implicit yet meaningful patterns within qualitative interview data.

Following the identification and classification of reasons, responses that addressed similar topics were grouped into broader thematic categories. This step enabled us to organise the data into a set of distinct reason types, each of which was then linked to the appropriate MHC category or categories.

\subsubsection{Inter-coder Reliability}
To ensure the reliability and credibility of the coding process, the authors of this paper independently applied the coding scheme to the interview data. Each coder reviewed the transcripts separately, identifying considerations, reasons, and their corresponding MHC categories. Following this initial round of independent coding, the researchers met to compare their results, discuss discrepancies, and reach consensus on any disagreements. This process not only enhanced the clarity and consistency of the category definitions but also helped to minimise individual coder bias.

We adopted a consensus-based approach to strengthen inter-coder reliability and ensure the trustworthiness of the analysis, in line with established practices in qualitative research. \citet{hsieh2005three} highlight the importance of investigator triangulation and collaborative coder discussions in validating interpretations during directed content analysis. Similarly, prior studies have demonstrated that consensus coding can reduce subjectivity and improve the credibility of category assignments \citep{hill2005consensual, campbell2013coding}.

\subsection{Results}
\label{2_results}

In response to Q2, Q3, and Q4, most experts described hypothetical traffic situations and outlined the considerations and actions that AVs should take before executing a manoeuvre in these contexts. They also provided explanations, articulating why the considerations they mentioned were important and why they believed AVs should exhibit particular behaviours.

Based on thematic analysis, we identified thirteen distinct categories of reasons used by experts to justify AV behaviour. These reasons encompass concerns related to legal compliance, fairness, cultural norms, safety, and more. Table~\ref{tab:reason-summary} summarises how these categories map onto different levels of AV behaviour—normative, strategic, tactical, and operational—based on how experts described the type and proximity of reasoning involved.

\subsubsection{Regulatory Compliance}
\label{1_regulatory_compliance}

Experts identified regulatory compliance as a key reason for AV behaviour. Several noted that adherence to traffic laws is important not only to satisfy legal requirements but also to reflect public expectations and provide a clear behavioural standard in ethically uncertain situations (ID~1, 3, 11, 12). This suggests that rule-following by AVs is not merely a legal necessity, but also a social function that fosters predictability and mutual understanding among road users. When AV behaviour is predictable, it reduces the likelihood of hesitation or misinterpretation—conditions that can otherwise escalate into ethically complex or unsafe scenarios.

This expectation extended beyond abstract principles. One expert described how AVs should behave like “good drivers”—that is, in ways that are interpreted by others as law-abiding (ID18). Another highlighted the importance of logging incidents such as red-light violations, which could support public accountability (ID16). These examples demonstrate how regulatory compliance can be embedded in both the design of AV behaviour and broader system functionality.

Overall, regulatory compliance was not framed narrowly as a legal obligation but rather as a broader social and design concern—closely tied to how AVs interact with human road users and how systems can be built to support transparency and accountability. This framing implicates multiple human agents across different behavioural levels. At the normative level, lawmakers define the legal parameters that AVs must follow. At the tactical level, road users interpret compliance through visible, socially recognisable behaviours such as lane discipline and consistent signalling. At the operational level, system designers are responsible for ensuring that these behaviours are encoded into AV control architectures.

\subsubsection{Social Legitimacy}
\label{2_social_legitimacy}

Experts also identified social legitimacy as a relevant reason for AV behaviour. Unlike regulatory compliance, which focuses on adherence to formal rules, this category reflects the idea that AVs should act in ways perceived as fair, courteous, and socially appropriate by other road users.

Several experts emphasised that compliance with traffic rules should not be limited to strict legal interpretation. Rather, AVs should behave in ways that align with how rules are understood and applied in everyday contexts. One expert noted that such social understanding is important for minimising risk—particularly in situations where human drivers might informally violate traffic rules due to differing interpretations, allowing the AV to respond more safely and appropriately (ID1). Another expert argued that AVs should behave in ways that match public expectations based on typical traffic scenarios. For example, stopping at a red light even when no immediate hazard is present was framed as a demonstration of integrity and respect for shared norms (ID6).

Predictable, socially recognisable behaviour was consistently viewed as essential for helping others interpret AV intentions and for supporting mutual understanding in mixed-traffic environments. Social legitimacy, therefore, serves as a critical element of trust-building and behavioural coordination between human and automated agents.

Other experts pointed to the role of politeness and cultural awareness in shaping socially legitimate AV behaviour. One expert noted that AVs should drive in ways that reflect courteous and respectful conduct, helping to support cooperative interactions in shared traffic spaces (ID7). Another emphasised the importance of cultural sensitivity, explaining that how people interpret ethical priorities—such as who should yield—can vary across contexts. For example, while cyclists in both the Netherlands and the United States may share a general commitment to avoiding harm, their behaviour at pedestrian crossings may differ, reflecting local norms rather than moral disagreement (ID8). These perspectives suggest that social legitimacy requires AVs to recognise not only universal ethical values, but also context-specific practices.

Perception also played a central role in how experts framed socially acceptable AV behaviour. One expert argued that AVs must avoid actions that appear threatening (ID~11). In this view, how AV behaviour is interpreted in real time can matter as much as the technical details of what the vehicle does—reinforcing the idea that social acceptance depends on interactional fluency as well as rule compliance.

Taken together, these responses suggest that social legitimacy depends on AVs behaving in ways that are intelligible, predictable, and acceptable within a shared human environment. This includes visible signs of courtesy, cultural awareness, and respectful conduct that reinforce public trust and social cohesion.

At the normative level, social legitimacy is shaped by the expectations of road users, the general public, and local communities (e.g., differing norms among cyclists in the Netherlands and the U.S.). At the tactical level, actors such as pedestrians, cyclists, and human drivers play a central role by interpreting AV behaviour in real time and adjusting their responses accordingly.

\begin{landscape}
\begin{table}[H]
\label{tab:reason-summary}
\centering
\scriptsize
\setlength{\tabcolsep}{3pt}
\begin{tabular}{p{4cm} p{5.125cm} p{5.125cm} p{5.125cm} p{5.125cm}}
\toprule
\textbf{Reason Category} &
\textbf{Moral / Normative} &
\textbf{Strategic} &
\textbf{Tactical} &
\textbf{Operational} \\
\midrule

\textbf{Regulatory Compliance} &
- Follow traffic rules as societal duty and moral baseline \textsuperscript{(L)} (ID~1,3,11,12) &
\emph{Not mentioned} &
- Behave like a predictable “good driver” \textsuperscript{(RU)} (ID~18) &
- Alert and log red-light running \textsuperscript{(D)} (ID~16) \\

\textbf{Social Legitimacy} &
- Visible compliance signals fairness, integrity, cultural respect \textsuperscript{(RU,GP,LC)} (ID~1,6,8) &
\emph{Not mentioned} &
- Courteous, non-threatening driving to gain trust \textsuperscript{(RU)} (ID~7,11) &
\emph{Not mentioned} \\

\textbf{Environmental Responsibility} &
- Reduce emissions for sustainability \textsuperscript{(GP, D)} (ID~17) &
- Smooth traffic flow to cut fuel / stop-and-go \textsuperscript{(D)} (ID~17) &
\emph{Not mentioned} &
\emph{Not mentioned} \\

\textbf{Fairness \& Equality} &
- Equal treatment \textsuperscript{(GP, RU)} \newline
- Protect VRUs and animals \textsuperscript{(GP, RU)} \newline 
- Designers bear safety duty 
\textsuperscript{(GP, RU)} (ID~9,10,13) &
- Avoid exclusionary AV-only lanes \textsuperscript{(D)} \newline 
- Ensure safety functions for all \textsuperscript{(D)} (ID~5,12) &
- Adjust driving to shield vulnerable users \textsuperscript{(VRU)} (ID~1) &
\emph{Not mentioned} \\

\textbf{Efficient \& Trip Optimisation} &
\emph{Not mentioned} &
- Balance speed with trip efficiency \textsuperscript{(D,O)} \newline
- Honour faster-arrival goals  \textsuperscript{(D,O)} (ID~1,6) &
- Avoid over-caution that disrupts flow \textsuperscript{(RU)} (ID~16) &
\emph{Not mentioned} \\

\textbf{User \& Public Acceptance} &
\emph{Not mentioned} &
- Ensure passenger safety and comfort for acceptance \textsuperscript{(O)} (ID~5) &
- Drive in ways passengers find comfortable and acceptable \textsuperscript{(O)} (ID~5) &
\emph{Not mentioned} \\

\textbf{Cultural \& Norm Adaptation} &
- Uphold legal standards despite unsafe local habits \textsuperscript{(L, LC)} (ID~17) &
- Adapt to region-specific traffic behaviours  \textsuperscript{(D, RU} (ID~8)  &
- Adapt yielding / right-of-way to local norms \textsuperscript{(RU)} (ID~8) &
\emph{Not mentioned} \\

\textbf{Risk Minimisation \& Safety Assurance} &
- Adapt speed if safer \textsuperscript{(D, GP)} (ID~17) &
- Limit speed in pedestrian zones \textsuperscript{(RU)} 
\newline - Minimise manoeuvres \textsuperscript{(RU)} 
\newline - Plan for sensor-failure contexts \textsuperscript{(RU)} (ID~4,13,15) &
- Safe overtake / merge \textsuperscript{(RU)} (ID~1) \newline - Anticipatory VRU buffers\textsuperscript{(VRU)} (ID~2) \newline 
- Early hazard signal and brake for violator\textsuperscript{(RU)} (ID~14,16) \newline - Extra buffer when visibility blocked \textsuperscript{(RU)} (ID~2) &
- Detect and signal sudden obstructions early \textsuperscript{(D)} (ID~14) \\

\textbf{Human Interaction Management} &
- Transparent communication upholds fairness \textsuperscript{(D, GP)} (ID~15) &
- Distinguish reactive vs.\ goal-driven actions \textsuperscript{(D)} (ID~13) &
- Detect users, infer intent, and signal manoeuvres \textsuperscript{(O)} (ID~3,9,15) \newline 
- Predict pedestrian motion \textsuperscript{(VRU)} (ID~4,7,8) \newline 
- Cooperative merge / influence traffic \textsuperscript{(RU)} (ID~7,18) \newline 
- Decelerate to signal crossing \textsuperscript{(RU)} (ID~11) &
\emph{Not mentioned} \\

\textbf{Comfort \& User Experience} &
\emph{Not mentioned} &
- Integrate safety, efficiency, comfort \textsuperscript{(D, O))} (ID~7) &
- Maintain comfort to avoid overrides \textsuperscript{(O)} (ID~5) \newline 
- Smooth merge / decel \textsuperscript{(O, RU)} (ID~13) \newline 
- Avoid harsh braking and needless yielding \textsuperscript{(O)} (ID~9) &
\emph{Not mentioned} \\

\textbf{Continuous Vehicle Control} &
\emph{Not mentioned} &
\emph{Not mentioned} &
\emph{Not mentioned} &
- Maintain continuous environmental monitoring \textsuperscript{(D)} (ID~14) \\

\textbf{Control Transition (Takeover Requests)} &
\emph{Not mentioned} &
\emph{Not mentioned} &
\emph{Not mentioned} &
- Give clear, timely takeover warning \textsuperscript{(D, O)} (ID~5) \\

\textbf{Driver / System Vigilance \& Readiness} &
\emph{Not mentioned} &
\emph{Not mentioned} &
\emph{Not mentioned} &
- Do not depend on continuous driver alertness \textsuperscript{(D, O)} \newline 
- Manage engagement \textsuperscript{(D, O)} (ID~14) \\

\bottomrule
\end{tabular}
\caption{Expert-elicited AV behaviour expectations across reason categories (rows) and behavioural proximity levels (columns). \emph{Not mentioned} indicates no corresponding expert response for that level. Parenthetical codes indicate the primary relevant agent(s): (L) = Lawmaker/Policymaker/Regulator (RU) = Road Users, (VRU) = Vulnerable Road Users, (D) = Designer, (O) = Occupant, (GP) = General Public, (LC) = Local Community}.
\end{table}    
\end{landscape}

\subsubsection{Environmental Responsibility}
\label{3_environmental_preservation}

One expert framed environmental responsibility as a reason for AV behaviour. The recommendation was that AVs should aim to minimise traffic disruption and emissions—not only to improve flow efficiency, but also to reduce environmental harm and road safety risks (ID~17). The justification reflects a dual rationale. At the strategic level, designers are seen as responsible for planning AV behaviour that promotes smoother traffic flow, thereby decreasing fuel consumption and avoiding inefficient stop-and-go driving. At the moral or normative level, both designers and members of the general public play a role: designers are expected to embed sustainability goals into AV systems, while the public provides broader ethical impetus by valuing environmental protection and public well-being.

This response illustrates how strategic planning and moral obligations intersect in AV design. The AV is expected to balance operational efficiency with a commitment to minimising environmental harm.

\subsubsection{Fairness and Equality}
\label{4_road_user_equality}

Fairness and equality emerged as key reasons for AV behaviour, with several experts emphasising the importance of protecting vulnerable individuals, preventing systemic bias, and ensuring equitable access to mobility. These concerns extended beyond abstract values, pointing to specific behaviours and design choices that AVs should embody.

One expert highlighted that AVs must treat all road users equally, regardless of financial status or social identity. For example, pedestrians and vehicle occupants should receive the same level of consideration, irrespective of vehicle ownership or access to public space (ID10). Building on this, another expert argued that safety-critical features should not be restricted to those who can afford premium systems, but made available to the wider public as a matter of equity (ID12). These views reflect expectations that AV technology must not reinforce socioeconomic disparities in how safety is distributed.

Infrastructure design was also identified as a determinant of fairness. One expert cautioned that AV-only lanes could marginalise those unable to afford automated vehicles, creating exclusionary spaces within the transport network (ID~5). Ensuring fairness, from this perspective, requires inclusive planning that does not restrict access based on economic status or technology ownership.

Beyond access, several responses addressed fairness in AV driving behaviour. One expert noted that AVs should adjust their driving in specific contexts—such as merging or navigating intersections—to protect both vehicle occupants and vulnerable road users (ID1). Another extended this moral concern to animal welfare, arguing that AVs have a duty to avoid harm to both people and animals in their environment (ID13). These examples place fairness within the tactical level of decision-making, where real-time choices must prioritise those at risk.

Finally, concerns around accountability and design responsibility also emerged. One expert argued that AVs should not rely on human users to uphold safety—particularly in situations where attention may lapse. Instead, designers should implement effective fallback mechanisms to ensure reliable performance even in the absence of user intervention (ID~9). This perspective positions system designers as the primary agents responsible for operationalising fairness, especially in contexts where failures would disproportionately affect those least able to influence the vehicle’s actions.

Taken together, these responses frame fairness and equality not simply as aspirational ideals, but as practical imperatives in AV design, behaviour, and policy. At the normative level, expectations around equal treatment are shaped by the public and broader societal values. At the strategic level, designers are responsible for ensuring that infrastructure and deployment models do not create exclusion or bias. At the tactical level, AVs themselves must adapt in real time to prioritise the safety and dignity of vulnerable road users and even animals, expressing fairness through their driving decisions.

\subsubsection{Efficiency and Trip Optimisation}
\label{5_time_efficiency}

Experts identified efficiency and trip optimisation as key reasons for AV behaviour, particularly in how vehicles plan over time and respond moment by moment. Some experts framed this as a planning issue. One described how AVs should balance safety with trip efficiency, noting that while safety is essential, prioritising zero risk could result in no movement at all (ID1). Another expert emphasised the importance of aligning AV decisions with user expectations—such as whether to overtake or prioritise a quicker return home—highlighting the need for systems to adapt route selection and driving style based on individual passenger preferences (ID6). These responses reflect a view of AVs as long-term planners, tasked with optimising both safety and the travel experience across the entire journey.

Other experts focused on moment-to-moment behaviour. One warned that overly cautious driving—such as unnecessary stops or excessive yielding—could disrupt traffic flow and frustrate other drivers (ID~16). Here, efficiency was framed not merely in terms of arrival time, but also in how smoothly AVs integrate into shared traffic environments. While the expert reaffirmed that safety is non-negotiable, they argued that AVs should avoid being so passive that they disrupt the expectations of other road users or cause avoidable delays.

In summary, these perspectives suggest that efficiency and trip optimisation operate across multiple levels of AV reasoning: from high-level route planning to real-time traffic decisions. AVs are expected to align with passenger goals while avoiding behaviours that hinder flow. They must plan with foresight, while responding flexibly to traffic dynamics. At the strategic level, designers are responsible for developing systems that balance safety with efficiency across the full journey. Users and passengers also contribute by setting preferences and expectations. At the tactical level, nearby human drivers interpret AV behaviour in real time and are directly affected by excessive caution or inefficiency.

\subsubsection{User and Public Acceptance}
\label{6_acceptance}

One expert emphasised that autonomous vehicles (AVs) must consider both safety and comfort as essential components for achieving user and public acceptance (ID~5). The justification involved designing AV behaviour that not only prevents harm but also feels acceptable over time. According to this expert, AVs should ensure the safety of both passengers and other road users in the surrounding environment. Safety was described as a non-negotiable requirement.

At the same time, the expert highlighted that passenger comfort and perceived safety play a critical role in acceptance. If passengers do not feel safe or comfortable with the AV’s behaviour, they may choose to intervene, which could lead to risky situations. This illustrates that passenger experience must be taken into account to support both safe operation and user trust.

The expert also noted that comfort in social interactions—such as smooth lane changes or natural merging—is an important consideration, even though safety should remain the highest priority. Taken together, these points reflect the view that for AVs to gain broad acceptance, they must behave in ways that are not only technically safe but also perceived as appropriate and comfortable by both passengers and other road users.

\subsubsection{Cultural and Norm Adaptation}
\label{7_cultural}

In addition to legal compliance, several experts emphasised the importance of AVs being able to adapt to local cultural norms and informal traffic practices. The challenge, they suggested, lies in responding appropriately to region-specific expectations without compromising legal or ethical standards.

One expert observed that acceptable driving behaviour can vary significantly across regions, highlighting the need for AVs to recognise and adapt to local conventions—such as customary yielding practices (ID8). A related comment underscored the influence of broader cultural expectations, noting that differences in traffic environments can result in divergent assumptions about right-of-way or lane usage (ID8). This form of adaptation was framed as both strategic and ethical: it enables AVs to interact more effectively with other road users while also demonstrating respect for locally shared norms and values.

At the same time, experts cautioned against uncritical conformity to unsafe local habits. One expert argued that even when human drivers routinely ignore traffic rules—such as at roundabouts—AVs should continue to follow legal standards (ID~17). From this perspective, cultural flexibility must not override core commitments to safety and lawful conduct.

These responses suggest that culturally adaptive behaviour is essential for fostering public trust and ensuring smooth traffic participation, but it must be balanced with a commitment to consistent ethical and legal norms. At the normative level, local communities define expectations for courteous conduct, while regulators are responsible for ensuring that such adaptations do not undermine legal or ethical integrity. At the strategic level, designers must account for regional traffic patterns—such as local yielding customs—without reinforcing unsafe practices. Road users also play a role at this level, as their collective behaviours shape the informal norms AVs are expected to interpret. At the tactical level, the AV system must enact these adaptations in real-time decision-making when interacting with human drivers, cyclists, and pedestrians.

\subsubsection{Risk Minimisation and Safety Assurance}
\label{8_risk}

Risk reduction and safety assurance were consistently emphasised across expert responses, particularly in situations involving uncertainty, vulnerable road users (VRUs), and rapidly changing traffic conditions. Many of these justifications centred on how AVs respond tactically to their immediate surroundings, though some also reflected broader moral and strategic concerns.

Several experts described how AVs should manage manoeuvres such as overtaking or merging in ways that minimise the risk of collision (ID~1). These decisions require real-time risk assessments and trade-offs. For example, staying behind a vehicle may initially pose little risk, but prolonged hesitation could increase the likelihood of unsafe conditions as traffic dynamics evolve.

Protecting vulnerable road users was another recurring theme. One expert recommended that AVs take anticipatory action even when the presence of a VRU is uncertain—for example, when a pedestrian or cyclist might enter an intersection, even if such behaviour is not legally anticipated (ID2). Another stressed the need to maintain extra buffer space when visibility is limited (ID2). Both responses highlight the importance of proactive adjustments in response to incomplete or ambiguous environmental cues.

Speed control and manoeuvre planning were also seen as central to safe AV behaviour. Experts suggested that AVs should limit their speed in pedestrian zones (ID4) and adjust their driving decisions based on the position of nearby road users and the AV’s own operational state (ID13). These views reinforce the importance of interpreting the driving environment in ways that support safe, context-aware decision-making.

Some responses addressed the early detection of hazards. One expert noted that AVs should detect and signal environmental changes at the earliest possible stage to prevent risk escalation (ID~14). This type of anticipatory awareness reflects expectations at the operational level of control.

Other experts discussed the need to balance safety with efficiency. AVs, they suggested, should avoid unnecessary manoeuvres that increase complexity or disrupt flow, while retaining the flexibility to adapt during situations such as sensor failure (ID~15). These perspectives integrate short-term responsiveness with long-term reasoning about trip continuity and system resilience.

Intersection-related risks were also noted. AVs should be able to detect red-light violations and brake in time to prevent collisions, especially when other road users behave unpredictably (ID~16). This highlights the importance of reliable threat recognition and fast, calibrated responses in complex environments.

One expert framed safety as a moral responsibility. They argued that AVs should adjust their behaviour to prevailing traffic speeds—even if this means diverging from posted speed limits—when rigid rule-following would increase risk (ID~17). This reflects a view of safety that balances legal compliance with context-sensitive ethical judgement.

Overall, these responses portray risk minimisation as a layered responsibility involving tactical decision-making, operational awareness, and moral reasoning. AVs are expected to manage risk in real time while aligning their actions with both societal safety expectations and the practical demands of dynamic traffic. At the moral or normative level, safety is framed as an ethical obligation. Society expects AVs to prioritise harm avoidance, even if this occasionally requires deviating from strict legal rules. Designers are responsible for embedding this principle into system objectives. At the strategic level, designers must ensure that AV systems can manage long-term risk factors—such as sensor failures or route planning decisions—without compromising safety or vehicle stability. At the tactical level, AVs are expected to make real-time driving decisions—such as overtaking, merging, speed regulation, and anticipating VRU behaviour—that maintain safe interactions in complex settings. At the operational level, the AV system must continuously monitor the environment and detect potential hazards early enough to allow effective, non-disruptive intervention.

\subsubsection{Human Interaction Management}
Experts highlighted that AVs must be capable of managing interactions with human road users by demonstrating social awareness, responsiveness, and clear communication. These expectations reflect the need for AVs to behave in ways that support mutual understanding, predictability, and coordination in shared traffic environments.

A recurring theme in expert responses was the importance of perceiving and anticipating the intentions of surrounding road users. One expert emphasised that AVs must detect nearby actors and infer their likely behaviour in order to respond appropriately in real time (ID3). Others noted that this type of tactical responsiveness is especially critical at intersections or unmarked crossings, where pedestrian movements or driver decisions may be uncertain or difficult to predict (IDs4, 7, 8).

In addition to perception and prediction, experts underlined the need for AVs to communicate their own intentions clearly. Several pointed to the value of signalling manoeuvres through visual cues or movement-based indicators, enabling pedestrians and other road users to interpret the AV’s next action (IDs3, 9, 15). In some cases, such communicative behaviour was framed not only as a practical requirement but also as a moral obligation—for example, using visual signals to alert others if a hazard arises (ID15).

Interaction was also described as bidirectional. AVs must not only respond to human behaviour but also influence it in cooperative ways. For instance, one expert explained that successful merging involves dynamically judging speed and spacing, so the AV’s behaviour encourages safe and predictable responses from others (IDs~7, 18). In this view, AVs are expected to actively shape traffic interactions rather than merely reacting to them.

Some responses addressed how AVs distinguish between reactive behaviour and navigation goals. One expert noted that it is important for AVs to determine whether a given action serves a short-term tactical response or contributes to a longer-term strategic objective. This distinction between tactical and strategic reasoning was seen as essential for maintaining safe, understandable interactions with human road users (ID~13).

Finally, movement itself was sometimes highlighted as a communicative tool. One expert observed that deceleration can signal to pedestrians that the AV is yielding or inviting them to cross, reinforcing non-verbal interaction as a safety-relevant behaviour (ID~11).

Taken together, these responses emphasise that effective human interaction management in AVs requires socially legible, adaptive, and cooperative behaviours. While rooted in short-term tactical reasoning, such behaviours also carry normative significance through their contribution to fairness, safety, and mutual understanding. Relevant human agents include pedestrians, cyclists, drivers, passengers, and others navigating shared environments alongside AVs.

\subsubsection{Comfort and User Experience}
Several experts noted that comfort in AV behaviour is not merely an aesthetic concern but a practical factor with implications for safety, social acceptance, and strategic planning. While most of these considerations operate at the tactical level, some also reflect longer-term moral and planning concerns.

One expert pointed out that AVs should ensure passenger comfort and stability to reduce the risk of user override or disengagement—situations that may occur when occupants feel anxious or uncertain (ID~5). Here, comfort is framed as a factor in system reliability, where the quality of the passenger experience contributes directly to ongoing safety and trust in the vehicle’s control.

Other responses emphasised that comfortable driving behaviour is important not only for passengers, but also for how other road users interpret AV actions. One expert noted that smooth merging and gradual deceleration help align AV behaviour with human expectations, avoiding sudden movements that could startle or confuse others (ID~13). This conduct was linked to moral expectations around fairness and predictability in shared public spaces.

Comfort was also described as an integrated design goal. One expert suggested that AVs should aim to meet the combined needs of safety, efficiency, and comfort—rather than treating them as separate or conflicting objectives (ID~7). In this view, AV decision-making must simultaneously address ride quality, ethical obligations, and trip performance.

Specific driving behaviours were mentioned as relevant to user experience. One expert recommended that AVs avoid harsh braking, which can degrade ride comfort and may be perceived as a loss of control (ID9). Another advised against unnecessary yielding when such actions could create confusion or instability for passengers—even while the AV remains attentive to the needs of other road users (ID9).

In summary, these responses show that managing comfort is essential to how AVs are perceived and trusted. Passenger well-being is not separate from system safety but a condition for it—influencing how users respond to automation and how AVs are judged by others in traffic. At the strategic level, designers are expected to integrate comfort alongside safety and efficiency as core planning objectives, while users and passengers play a role in defining expectations. At the tactical level, the AV system itself must implement comfort-sensitive behaviours that contribute to a positive passenger experience and influence public interpretation of AV actions.

\subsubsection{Continuous Vehicle Control}
Attention to ongoing traffic conditions was described as an essential feature of AV behaviour, even during routine or seemingly low-risk segments of a journey. One expert highlighted the importance of maintaining continuous awareness of the driving environment to ensure that AVs can respond effectively to sudden or unexpected changes (ID~14). The justification for this expectation was grounded in the need for real-time perception and immediate responsiveness—capabilities that must be sustained throughout the entire trip, not just in high-risk moments.

This was framed as a design requirement, rather than a situational enhancement. The expert emphasised that operational vigilance must be a core component of everyday AV functionality, enabling the system to monitor its surroundings and make timely adjustments without external prompting.

In conventional vehicles, this function is typically carried out by the human driver, whose attentiveness serves as the standard for responsive control. In AVs, however, the system itself bears this operational responsibility.

Taken together, the response points to the need for persistent, low-level monitoring as a baseline expectation of AV behaviour. Relevant human agents in this context includeing 1.) The AV system, which must continuously perceive and adapt to environmental inputs; and 2.) The human driver, who serves as a reference point for acceptable vigilance and responsiveness in real-world driving.

\subsubsection{Driver/System Vigilance and Readiness}
A concern raised by one expert focused on the risks of assuming that human drivers can remain continuously alert during extended periods of semi-automated driving (ID~14). The response highlighted the natural decline of human attention in passive supervision roles, where drivers are expected to intervene if needed but are not actively engaged in the driving task. Designing AV systems that rely on constant human fallback in such scenarios introduces potential safety hazards.

Rather than assuming perfect vigilance, the expert suggested that AVs must account for human cognitive limitations through system-level design. This includes either reducing dependence on driver intervention altogether or incorporating features that monitor and manage driver engagement in real time. The expectation is that AV systems should reflect a realistic understanding of the human role during shared-control operation, ensuring safety even when driver readiness cannot be guaranteed.

The relevant human agent in this context is the driver, whose attentional limitations must be anticipated and supported through the system’s operational logic and interface design.

\subsubsection{Control Transition (Takeover Requests)}
In contexts where automated systems still rely on human oversight, one expert emphasised the importance of providing drivers with adequate warning before control is handed back to them (ID~5). The aim is to ensure that drivers have sufficient time to regain situational awareness and resume control without confusion or hesitation.

This concern is particularly relevant to partially automated systems—such as Level 2 and Level 3 vehicles—where the AV handles driving functions under normal conditions but expects the human driver to take over when prompted. According to the expert, these transitions must be carefully managed and clearly communicated, as they represent moments of elevated risk.

The underlying justification relates to the system’s ability to accommodate human attention and reaction time. Even brief delays or ambiguous cues during handover can compromise safety. As such, AVs must be designed to anticipate these limitations, ensuring that takeover procedures are both timely and predictable.

Here, the relevant human agent is the driver, whose ability to respond safely depends on how effectively the system facilitates the transition. This highlights an operational requirement within AV design, in which system behaviour must be aligned with human cognitive constraints during real-time decision-making.

\subsection{Discussion}

To address Objective 1, we aimed to identify and structure the types of human reasons that should inform AV manoeuvre planning. Our study contributes to addressing the \textbf{practical gap} by offering a layered mapping of thirteen reason categories, organised across moral, strategic, tactical, and operational levels, and explicitly linked to the roles of relevant human agents. This structure transforms abstract ethical and functional expectations into actionable guidance for AV decision-making, in alignment with the tracking principle of Meaningful Human Control (MHC).

This section also responds to the \textbf{methodological gap} identified in the introduction by demonstrating how expert interviews and inductive analysis can uncover the structure of human reasons relevant to AV decision-making. Our approach systematically captured how experts from diverse domains interpret what matters in AV behaviour. Rather than imposing pre-existing classifications \citep{calvert2020conceptual}, we allowed reason categories and their associated behavioural layers to emerge inductively. The resulting framework (Table~\ref{tab:reason-summary}) connects ethical concerns with system architecture, mapping reason types to behavioural levels (from normative to operational) and assigning responsibility to relevant human agents such as lawmakers, designers, occupants, and road users. This structure provides a practical bridge between high-level values and real-world implementation.

\vspace{0.75em}
\noindent\textbf{Multiple overlapping reasons in a single manoeuvre.}
A central insight from our findings is that AV manoeuvre planning rarely relies on a single type of reason. Instead, a single manoeuvre often engages multiple overlapping reasons that reflect different layers of ethical and practical concern. For example, Expert ID~17 connected environmental responsibility, cultural adaptation, and risk minimisation—demonstrating that a single driving action (e.g., cautious merging) might be justified by ecological sustainability, cultural sensitivity, and hazard avoidance simultaneously. Additionally, reason categories themselves are not confined to one behavioural layer. Regulatory compliance, for instance, spans normative expectations (e.g., respecting laws), tactical execution (e.g., behaving like a predictable driver), and operational functionality (e.g., logging red-light violations). This layered nature of reasons suggests that manoeuvre planning systems must support multi-reason, multi-level responsiveness, rather than relying on rule-based execution alone.

\vspace{0.75em}
\noindent\textbf{Human proximity and agent roles.}
Another key finding concerns the relationship between reason type and the proximity of human agents. Moral and normative reasons were typically associated with more socially and institutionally distant agents—such as policymakers, the general public, and local communities—who define broad ethical standards. In contrast, tactical and operational reasons were more closely tied to agents physically or functionally proximate to AV activity, such as road users, vulnerable individuals, and vehicle occupants. Designers notably appear across all levels, from embedding values into the system at a normative level to implementing control features at the operational level. This supports and extends the proximity-based reasoning model introduced by \citet{mecacci2020meaningful} and elaborated by \citet{calvert2020conceptual}, which posits that meaningful control depends on responsiveness to human reasons distributed across different behavioural and social layers. Our findings provide empirical grounding for this framework and offer a structured elaboration of how proximity and agent responsibility are linked in AV design.

\vspace{0.75em}
\noindent\textbf{Variation in behavioural level depending on task interpretation.}
We also found that the behavioural level at which a reason is situated can vary depending on how the AV task is interpreted. For example, time efficiency is often treated as a strategic concern, but depending on the situation, it may also appear at tactical or even operational levels. A strategic interpretation might involve planning the most efficient route, while a tactical one may involve decisions such as overtaking or avoiding hesitation that disrupts flow. Despite being motivated by the same underlying reason—efficiency—these interpretations correspond to different layers of action. This highlights the importance of distinguishing between the justification for a behaviour and the specific behavioural level at which it is operationalised.

\vspace{0.75em}
\noindent\textbf{Changes across levels of automation.}
We also observed that the relevance of certain reasons shifts with the AV’s level of automation. For instance, considerations such as Control Transition and Driver/System Vigilance were particularly prominent for lower levels of automation (L2/L3), where human fallback is still required. At higher levels (L4/L5), these concerns recede, and the focus shifts towards trade-offs among values such as fairness, efficiency, and comfort—especially when AVs operate without direct human supervision. Our framework accommodates these shifts by revealing which reasons—and which agents—are most relevant at each automation stage and behavioural level.

\vspace{0.75em}
\noindent\textbf{Interpretative flexibility in core categories.}
A further nuance in our data involves the diverse interpretations of regulatory compliance. While several experts (e.g., IDs1, 3, 11, 12) framed rule-following as a strict moral baseline, others (e.g., ID17) viewed traffic laws as flexible guidelines to be overridden when necessary to ensure fairness or safety. This reflects the contextual and scenario-sensitive nature of AV ethics: manoeuvre planning must not only track rules but also balance them against competing values such as legitimacy, risk minimisation, and social understanding.

\vspace{0.75em}
\noindent\textbf{Positioning within existing literature.}
Our framework also offers a way to contextualise and relate existing efforts to define what AVs should consider when making driving decisions. Prior work has provided focused contributions on specific types of consideration: for example, \citet{UNECE2023, olleja2025validation} define legal and behavioural benchmarks for competent driving; \citet{geisslinger2021autonomous} formalise ethical principles for risk-sensitive planning; \citet{schwarting2018planning} model social value orientation for cooperative behaviour; and \citet{thornton2018value} apply value-sensitive design to embed stakeholder values in AV system logic. While these approaches differ in their aims and methodologies, our framework does not attempt to replace or rank them. Rather, it provides a layered structure through which these contributions can be situated—by connecting the types of reasons they represent (e.g., legality, fairness, efficiency, comfort) to specific levels of AV behaviour (e.g., moral/normative, strategic, tactical, operational). In this sense, our empirically derived categorisation can serve as a common referential model—one that helps clarify how diverse AV design goals and values interact across system layers and in relation to different human agents.

\vspace{0.75em}
\noindent\textbf{Practical design guidance and limitations.}
Beyond theoretical insights, our structured mapping offers practical guidance for AV developers and policy designers. For example, developers working on L2/L3 vehicles may use our findings to prioritise clear and timely takeover cues, while those building L4/L5 systems may focus on fairness–efficiency trade-offs and behaviour intelligibility in mixed traffic. This structured mapping of thirteen reason categories across behavioural levels and agent roles can help translate ethical expectations into system-level specifications—by clarifying what kinds of human concerns should be considered, at which layer of system behaviour, and by whom. Furthermore, as the questions did not solely focus on ethically challenging situations, the identified reasons are applicable not only in edge cases but also in routine and general AV driving scenarios where aligning AV behaviour with human reasons is essential.

Nonetheless, our approach has limitations. First, the scope of this study is restricted to manoeuvre-level decision-making, rather than broader systemic influences such as infrastructure, corporate strategy, or legal frameworks. Second, interpretations of the identified reasons may vary across cultural and regional contexts, which could affect the generalisability of the findings. Our expert pool was primarily composed of individuals from the Netherlands, the United States, and the United Kingdom, with some representation from other technologically advanced countries, including Germany, Japan, and China. It is also important to note that the mapping of reason categories to behavioural layers should not be interpreted as exhaustive. In particular, cells marked “Not mentioned” in Table~\ref{tab:reason-summary} do not imply that no relevant reasoning exists at that level. Theoretical frameworks suggest that any reason could, in principle, be interpreted across multiple control layers. However, because our study is empirical in nature, the absence of content in certain cells reflects the limits of what was raised by experts, not a conceptual impossibility. Future research could further investigate these gaps through targeted questioning or normative modelling. Future work may also extend this approach by incorporating a broader and more diverse stakeholder base—such as regulators, insurers, and urban planners—or by developing prioritisation models to resolve conflicts among overlapping reasons.

\vspace{0.75em}
\noindent\textbf{Summary of contributions.}
In summary, addressing Objective 1, we identified thirteen categories of human reasons relevant to AV manoeuvre planning and mapped them to behavioural layers and responsible agents. Our findings highlight the complexity and ethical weight of even routine AV decisions and demonstrate that manoeuvre planning must respond to layered, sometimes conflicting, human expectations. By integrating these findings with the MHC framework, our study offers a pathway towards AV systems that behave in ways aligned with human reasons.

\section{A Principled Conceptual Framework for AV Behaviour in Ethically Ambiguous Routine Situations}
\subsection{Scenario and Questionnaire Design}
\label{sec:scenario_questionnaire_design}

To support the development of a principled conceptual framework for AV behaviour in ethically ambiguous routine situations, we designed a structured questionnaire centred on a specific overtaking scenario. This scenario involved an AV travelling behind a slow-moving cyclist on a two-way road marked with a double yellow line. This road marking typically prohibits overtaking, thereby introducing a regulatory constraint that renders the situation ethically ambiguous: the AV could either remain behind the cyclist at a reduced speed or initiate an overtaking manoeuvre by crossing the double yellow line.

After responding to Questions 2–4—which explored their reasoning about what the AV should consider when planning its manoeuvre—experts were shown a short video clip depicting the overtaking scenario (see Figure~\ref{fig:video_scenario}). They were then asked a series of structured follow-up questions (Questions 5–9), which invited them to predict how the AV would behave and to explain the reasoning behind their predictions. These questions were designed to elicit the experts’ interpretations of AV decision-making without explicitly prompting normative judgments about what the AV should do.

\subsection{Mapping Expert Reasoning and Framework Development}
\label{sec:mapping_reasoning_framework}

\textit{Qualitative Content Analysis (QCA)} \citep{schreier2012qualitative} was employed to analyse expert responses to Questions 5–9. These questions asked experts to predict what the AV would do in a specific overtaking scenario and to justify their predictions. Following Schreier’s structured approach, we developed a mixed-category coding strategy that incorporated both:

\begin{itemize}
    \item \textbf{Concept-driven categories}, used to group expert responses by predicted AV behaviour: either overtaking the cyclist or not overtaking the cyclist.
    \item \textbf{Data-driven categories}, used to inductively identify the specific justifications experts provided for their predictions.
\end{itemize}

The first stage of the analysis involved classifying each expert response according to the predicted action (i.e., follow vs overtake). Within each group, expert justifications were then collected and analysed to identify recurring patterns of reasoning. These inductively derived justifications were subsequently mapped to the predefined set of thirteen reason types originally developed from the open-ended responses to Questions 2–4 (see Section~\ref{2_results}).

As part of the interpretation phase, we conducted a qualitative synthesis of the coded data to explore patterns in how experts linked specific reasons to predicted AV behaviours. This involved examining which reasons frequently co-occurred, how the same reasons were interpreted differently across contexts, and instances where multiple reasons appeared within a single justification. This helped us understand how reasons relate to each other or are prioritised. The synthesis was guided by principles of thematic pattern analysis \citep{braun2006using} and constructivist grounded theory \citep{charmaz2014constructing}, with analytical interpretations discussed collaboratively between authors to enhance transparency and reduce individual bias.

\subsection{Results}
This section presents the findings from the expert interviews. Based on their responses to Q5, most experts interpreted the AV as the sole traffic participant whose actions were being evaluated. Two primary behaviours that the AV might adopt in the given scenario were identified: 
(1) \textit{following the cyclist}, and (2) \textit{overtaking the cyclist}.

In addition, some experts distinguished between what the AV is likely to do (\textbf{expected action}) and what the AV ought to do (\textbf{preferred action}). To maintain clarity, this distinction will be upheld throughout the remainder of the paper: expected action refers to what the AV is predicted to do, whereas preferred action refers to what the AV should do from the expert's normative perspective.

\begin{figure}[htbp]
    \centering
    \includegraphics[width=1.0\textwidth]{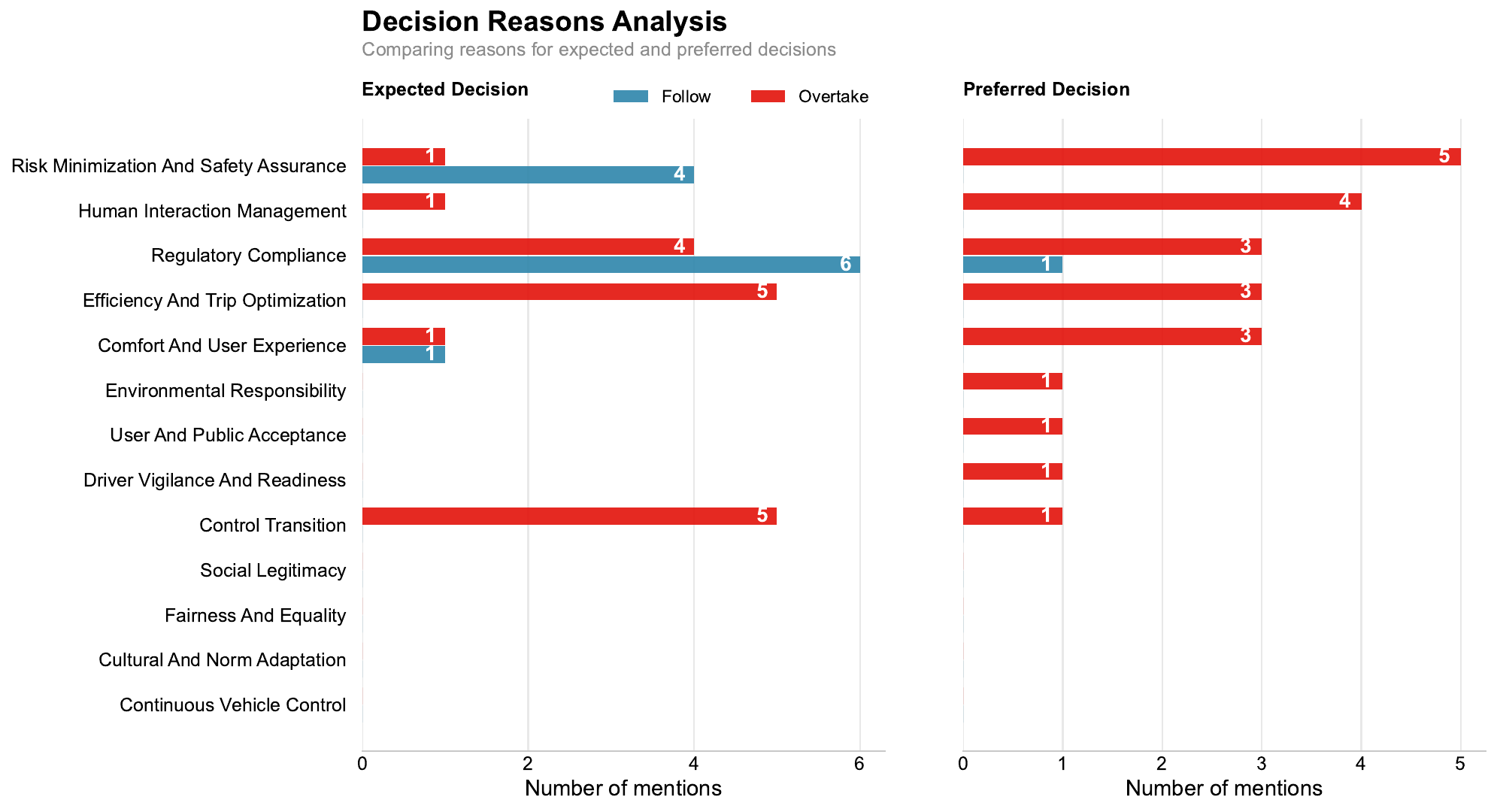}
    \caption{Experts’ reasons for predicting whether the AV will or should follow (a) or overtake (b) the cyclist. Thirteen reasons are listed in the first column, expert IDs in the second, and total counts in the third. Blue indicates predicted behavior (will), red indicates preferred behavior (should).}
    \label{fig:comparison_will_should}
\end{figure}

Figure~\ref{fig:comparison_will_should} summarises the reasons experts provided for predicting whether the AV \textbf{will} or \textbf{should} follow or overtake the cyclist. Thirteen distinct reasons are presented, grouped by action type. Blue shading represents predicted (``will'') actions, while lighter red shading represents preferred (``should'') actions.

In general, experts cited a more limited set of reasons for why the AV will follow the cyclist—most commonly grounded in \textit{regulatory compliance} and \textit{risk minimisation and safety assurance}. By contrast, overtaking was associated with a broader range of justifications—including \textit{efficiency and trip optimisation}, \textit{comfort and user experience}, and \textit{human interaction management}, alongside the aforementioned safety and legal concerns.

This pattern suggests that \textit{following the cyclist} is predominantly justified by a narrow range of risk-averse or rule-based considerations, whereas \textit{overtaking} is viewed as a more complex decision that draws upon a wider set of overlapping reasons. The following subsections present detailed results, including representative quotations, to explain the reasoning experts used in predicting whether the AV will or should follow or overtake the cyclist.

\subsubsection{Reasons for the AV Will Follow the Cyclist}
From Figure~\ref{fig:comparison_will_should}, panel~(a), the most frequently cited reasons for why the AV is expected to follow the cyclist were \textit{regulatory compliance} and \textit{risk minimisation and safety assurance}. Several experts (e.g., ID2, ID12) emphasised that the AV is programmed to obey traffic laws—such as not crossing a double yellow line—making overtaking legally impermissible. Others (e.g., ID3, ID11) pointed to the importance of safety, arguing that the AV would stay behind the cyclist to avoid potential collisions or unsafe manoeuvres. In many cases, both legal and safety considerations were closely intertwined in the experts' reasoning. One expert (ID2) also mentioned driver discomfort as a potential outcome, noting that while the AV is obligated to follow the cyclist, this may result in frustration or discomfort for the human passenger. However, this was framed not as a reason for the AV's behaviour, but rather as a consequence of its strict adherence to rules and safety protocols.

\subsubsection{Reasons for the AV Should Follow the Cyclist}
Notably, only one expert (ID12) explicitly stated that the AV \textit{should} follow the cyclist, as indicated by the lighter red shading in panel~(b). This expert argued that, in addition to the fact that current AVs are programmed to follow traffic rules, they \textit{should} continue to be programmed in accordance with those rules. Interestingly, this expert was the only one with a background in the legal aspects of AVs.

\subsubsection{Reasons for the AV Will Overtake the Cyclist}
In panel~(a) of the figure, the experts cited a variety of reasons for why the AV is expected to overtake the cyclist. The most frequently mentioned reasons included \textit{efficiency and trip optimisation}, the possibility of \textit{control transition}, and \textit{regulatory compliance}. \textit{Risk minimisation and safety assurance} and \textit{comfort and user experience} were discussed less frequently but still featured in some responses.

Several experts emphasised time efficiency as a key factor. One expert (ID15) stated that the driver would likely take over control and overtake the cyclist in order to arrive at work on time. Another expert (ID10) also mentioned the urgency of reaching a destination as a reason for manual takeover. A third expert (ID7) explained that while the AV might initially follow the cyclist, the driver would likely take over if delays occurred, particularly because they wouldn’t want to be late. Similarly, an expert (ID13) noted that overtaking would allow the AV to maintain a more efficient speed, and another (ID14) emphasised that roads are designed for higher speeds than cyclists typically travel, making it uncomfortable for drivers to go significantly slower than expected.

\textit{Regulatory compliance} was discussed in relation to whether overtaking is legally permissible. One expert (ID13) stated that in the UK, it is legal to overtake a cyclist if they are travelling below 10 mph, and therefore expected the AV to do so. Another expert (ID6) viewed the AV’s failure to overtake when the opposite lane is empty as irrational, suggesting that such behaviour would seem “stupid” to human drivers. An expert (ID7) commented that while drivers are aware of traffic rules, they often weigh the risk of breaking those rules against the need to stay on schedule. One expert (ID15) believed that the AV would not overtake because it is programmed to strictly follow traffic rules, but that the driver would override this behaviour when time becomes a priority.

The possibility of \textit{control transition} was also commonly mentioned. Multiple experts (IDs 6, 7, 10, 14, 15) described scenarios in which the human driver would take over control to overtake the cyclist. Some (e.g., ID7, ID10) mentioned this as a temporary takeover, while others (e.g., ID14, ID15) connected it to the frustration caused by prolonged low-speed travel. One expert (ID6) stressed that such a manoeuvre would not compromise safety and therefore believed the driver would proceed with overtaking.

A few experts raised \textit{social and comfort}-related concerns. One expert (ID13) emphasised the importance of not frustrating other road users who may be following the AV. Another expert (ID14) pointed out that drivers are not accustomed to travelling so slowly, especially on roads designed for higher speeds, and would likely feel discomfort in such situations—prompting a takeover.

Although less frequently mentioned, \textit{safety} still featured in the discussion. One expert (ID6) stated that overtaking would be acceptable if the opposite lane were empty, implying that safety would not be compromised. Another (ID10) said the driver would overtake only if it could be done without putting anyone at risk, suggesting that manual takeovers are still constrained by a concern for \textit{harm avoidance}.

\subsubsection{Reasons for the AV Should Overtake the Cyclist}

Several experts argued that the AV \textit{should} overtake the cyclist, even if it involves crossing a double yellow line, because failing to do so could cause confusion, frustration, and even safety issues. A common theme among the responses was that not overtaking could disrupt traffic flow, potentially leading to cascading negative effects—such as increased risk (ID1, ID17), discomfort (ID5, ID11), higher emissions (ID17), and reduced public acceptance (ID11). These disruptions were framed not only as inefficiencies but also as factors that could compromise safety and overall system performance.

For some experts, \textit{risk minimisation and safety assurance} did not necessarily align with strict rule-following. Instead, safety was interpreted contextually—such as maintaining adequate distance from the cyclist (ID8), supporting a steady traffic flow (ID16, ID17), or reducing cognitive workload for drivers (ID5). This suggests that pragmatic, situational decisions—even if technically in violation of traffic regulations—can still serve safety-related objectives.

Experts also emphasised that a single justification was rarely sufficient; rather, multiple factors often combined within a single rationale. For example, an expert might state that the AV should overtake because it is both \textit{safer} and \textit{more comfortable} (ID5), or because it improves \textit{traffic flow} and aligns with \textit{human driving behaviour} (ID17). These reasons were presented as equally important, rather than hierarchically ordered.

Regarding \textit{regulatory compliance}, experts acknowledged that overtaking may violate traffic rules (ID8, ID11, ID16). However, they generally agreed that strict rule adherence should not override other practical concerns such as \textit{safety} and \textit{traffic efficiency}. In this context, rule violations were often framed as acceptable when they led to better outcomes for all road users.

In addition, \textit{human interaction management} and \textit{comfort and user experience} played significant roles in shaping expert expectations. Some experts noted that cyclists may feel uncomfortable or stressed when a vehicle follows too closely without overtaking (ID5, ID11, ID17), and that passengers or drivers may become frustrated by overly cautious AV behaviour (ID1, ID11). These concerns link comfort with public trust and acceptance, suggesting that AVs should behave in ways that are intelligible and relatable to human road users.

Finally, some experts (e.g., ID5) stated that in such situations, they would personally choose to overtake the cyclist by taking control of the vehicle. This reinforces the view that manual takeover may remain a practical necessity when AVs are constrained by rules that fail to account for situational flexibility.

\subsection{Discussion}

To address Objective 2, we applied the reason framework developed in Section 3 to a context-specific case study involving a routine but ethically ambiguous AV scenario: overtaking a cyclist. This application contributes to \textbf{the theoretical gap} by operationalising the tracking condition of Meaningful Human Control (MHC) in everyday driving contexts, moving beyond high-stakes dilemmas to examine how AVs might align with human reasons in complex, real-world situations.

This section also speaks to \textbf{the methodological gap} by demonstrating how structured qualitative analysis—linking expert-elicited reasons to specific behavioural recommendations—can yield actionable ethical insights. By capturing expert judgments on both expected (“will”) and preferred (“should”) AV behaviours, we identify how contextual factors, value tensions, and individual reasoning strategies shape nuanced expectations for AV decision-making. These insights form the basis for the principled framework proposed in this section.

\begin{figure}
\includegraphics[width=\textwidth]{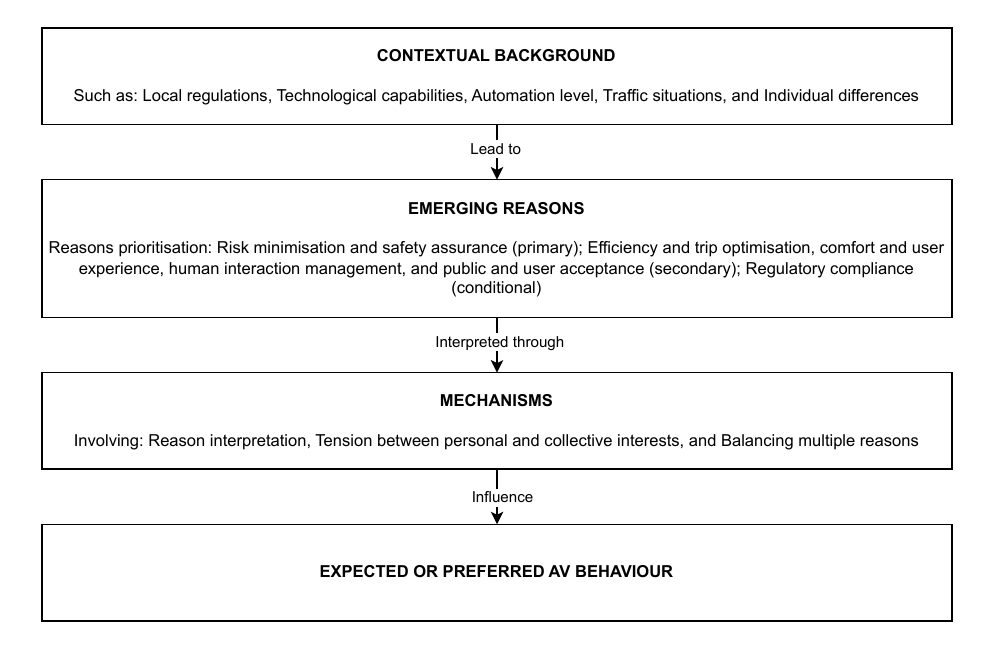}
\caption{Framework illustrating how human reasons influence the expected or preferred behaviour of AVs in ethically ambiguous routine situations. The framework highlights the role of contextual background in shaping these reasons and their impact on expected AV behaviour.}
\label{fig:reasons_influence_AV_behaviour}
\end{figure}

This focus on routine, ethically ambiguous scenarios expands on prior work that has largely centred on high-stakes dilemmas, such as crash scenarios or trolley problems \citep{rhim2021deeper, milford2025all}. Whereas those studies highlight binary moral choices, our study surfaces the nuance of everyday trade-offs and expert reasoning in context-rich decisions. This discussion synthesises those insights into a cohesive framework (Figure~\ref{fig:reasons_influence_AV_behaviour}) and considers their implications for AV development and policy.

\vspace{0.75em}
\noindent\textbf{Contextual Background Influencing Emerging Reasons.}
Our proposed framework begins with the influence of contextual background, which includes five critical factors: local regulations, technological capabilities, automation level, traffic situations, and individual differences. While the tracking condition in MHC theory requires that automated systems respond to human moral reasons \citep{santoni2018meaningful}, existing literature does not explain how such reasons emerge from contextual circumstances. This study contributes a novel insight by showing that expert reasons are shaped by situational background assumptions.

For instance, local regulations play a foundational role. One expert (ID13) stated that overtaking would be permissible under UK traffic laws. Conversely, experts ID7 and ID15 assumed stricter rules prohibiting overtaking, leading them to suggest manual takeover as a necessity. Technological capabilities and automation level also constrain or enable reasoning. Experts (IDs4, 15) noted that current AVs are programmed to avoid rule violations, thus requiring human intervention when overtaking is contextually necessary. This aligns with assumptions about Level~2 automation, where driver readiness remains essential. When experts assumed no manual override was available, they introduced alternative reasons such as \textit{environmental responsibility}, \textit{human interaction management}, and \textit{user and public acceptance}.

Traffic situations—such as encountering a slow cyclist on a bidirectional road—influenced reasoning around safety, vigilance, and comfort. For instance, Expert ID5 emphasised that prolonged following increases driver workload, reinforcing the need for AVs to act pragmatically. Individual differences in expert values were also significant. Expert ID11 advocated strict rule-following as a matter of principle, while Expert ID~6 endorsed a consequentialist view, suggesting AVs should act based on outcomes (e.g., safety), regardless of legality.

\vspace{0.75em}
\noindent\textbf{Distinct Reasons Leading to Different Expected Behaviors.}
Distinct sets of reasons translate into differing AV behaviors. As shown in Figure~\ref{fig:comparison_will_should}, experts cited \textit{regulatory compliance} and \textit{risk minimisation and safety assurance} as dominant reasons for why AVs are expected to follow the cyclist. In contrast, overtaking behavior was associated with a wider range of reasons, including \textit{efficiency and trip optimisation}, \textit{comfort and user experience}, \textit{user and public acceptance}, and \textit{human interaction management}.

Three mechanisms were identified as key to understanding how reasons lead to behaviors. First, reason interpretation varies by context. \textit{Risk minimisation and safety assurance}, for example, was interpreted as a reason to follow the cyclist (avoiding risky manoeuvres, ID4), but also to overtake (reducing traffic disruptions, ID17). This shows that a single reason can support opposing behaviors depending on situational framing.

Second, a tension between personal and collective interests was apparent. Personal motivations (e.g., arriving on time, IDs7, 15) often shaped expected behavior. Meanwhile, collective values (e.g., environmental sustainability, ID17; shared road safety, ID~5) influenced preferred actions. This tension reflects a broader ethical divide in AV decision-making, as noted in prior work \citep{bonnefon2016social}.

Third, balancing multiple reasons emerged frequently. Experts combined several motivations in a single rationale (e.g., ID~6 cited both safety and comfort). Notably, when regulatory compliance conflicted with more practical or safety-oriented reasons, the latter were often prioritised. These dynamic prioritisation mechanisms echo the tracking model proposed by \citet{mecacci2020meaningful}, in which AV systems are designed to respond to the most proximal reasons unless overridden by more distal values. Our findings complement this by empirically showing that human experts interpret and prioritise reasons fluidly, often allowing situational context to shape whether a reason like safety or legality dominates. This underscores the challenge of implementing fixed hierarchies of reasons in AV design and supports the need for flexible, context-sensitive reason-tracking mechanisms.

\vspace{0.75em}
\noindent\textbf{Implicit Prioritization Among Reasons.}
While experts cited distinct reasons for AV behavior, a clear prioritisation pattern emerged across interviews. \textit{Risk minimisation and safety assurance} consistently served as the primary justification, regarded as non-negotiable in both “will” and “should” judgments. In contrast, \textit{efficiency and trip optimisation}, \textit{comfort and user experience}, \textit{human interaction management}, and \textit{user and public acceptance} appeared frequently but were typically mentioned alongside safety, making them secondary. \textit{Regulatory compliance} was treated as conditional—important when it does not conflict with other goals, but often overridden when it did, especially by pragmatic considerations such as traffic flow or situational safety. Supporting reasons such as \textit{environmental responsibility}, \textit{driver vigilance and readiness}, and \textit{continuous vehicle control} were mentioned by fewer experts and typically in specific contexts, highlighting how situational factors and individual expertise shape the salience of different considerations.

This prioritisation structure aligns with previous findings in AV ethics literature, where safety is widely regarded as the paramount consideration. For instance, the Moral Machine experiment \citep{awad2018moral} and expert-based studies \citep{milford2025all} show broad consensus that risk minimisation should guide AV behavior. Similarly, regulatory compliance has been treated as a conditional obligation in earlier work, particularly when rigid adherence may compromise safety or efficiency \citep{ma2024analysing}.

However, our results extend this discussion by demonstrating that such prioritisation patterns are not limited to high-stakes dilemmas but also emerge in routine, ethically ambiguous situations. This suggests that value trade-offs are not confined to emergencies, but are an ongoing feature of everyday AV operations. Moreover, while existing frameworks often rely on normative claims or abstract models of ethical reasoning, our findings reveal how experts actually balance and contextualise competing considerations—including legality, comfort, and environmental responsibility—based on real-world constraints and assumptions. This contributes a more fine-grained, empirically grounded understanding of how prioritisation unfolds across different layers of AV behavior, and how certain reasons rise or recede in relevance depending on the driving context.

\vspace{0.75em}
\noindent\textbf{Proposed Principled Framework for AV Decision-Making.}
Building on these prioritisation patterns, we propose a conceptual framework for AV decision-making in ethically ambiguous routine situations. The framework synthesises expert reasoning into three core guidelines. First, AVs must always prioritise safety. Second, in contexts where strict legal compliance may inhibit legitimate values like comfort, efficiency, or environmental quality, AVs should favour pragmatic actions—provided that safety is fully maintained. Third, while legal compliance should remain the default behaviour, carefully constrained deviations may be justifiable when rule-following contradicts higher-priority values. This framework translates expert ethical intuition into actionable design principles for dynamic AV decision-making in complex, real-world scenarios.

Unlike previous models that focus on rare, high-stakes scenarios—such as the MIT Moral Machine’s global crash dilemma study \citep{awad2018moral}, or algorithmic approaches like Augmented Utilitarianism (AU) \citep{gros2025methodology}—our framework addresses a less examined but highly relevant domain: ethically ambiguous, everyday driving situations. The Moral Machine revealed diverse cultural preferences in life-or-death crash dilemmas, underscoring the challenge of creating globally acceptable AV ethics. However, its emphasis on binary, extreme scenarios limits its applicability to nuanced real-world contexts.

Similarly, AU advances ethical reasoning by incorporating diverse moral theories into adaptable goal functions grounded in empirical data. It uses attributes like harm, fairness, and legality, refined through participatory methods, to compute ethical decisions dynamically. While AU aims for a transparent, mathematically structured foundation for AV ethics, it remains focused on generalisable, formal ethical prioritisation, often in critical or rare scenarios.

In contrast, our framework complements these by providing a practice-oriented structure for dynamic, routine AV behaviour, grounded in expert judgment rather than abstract moral theory or crowdsourced moral preferences. By emphasising safety, pragmatic flexibility, and context-specific legal reasoning, we propose a system that captures complex everyday value tensions—like comfort or efficiency versus strict rule-following—common in everyday AV operation. Our work thus fills a gap by offering actionable guidance for routine, ethically challenging decisions that do not involve stark life-or-death trade-offs, but are nonetheless ethically meaningful and crucial for AV public trust and usability.

\vspace{0.75em}
\noindent\textbf{Implications and Recommendations.}
Our findings suggest that AV systems should incorporate flexible, context-aware decision logic. Developers should embed mechanisms to interpret reasons dynamically and prioritise them based on real-time conditions. Policymakers should consider enabling AVs to operate within regulated bounds that allow principled flexibility—particularly in routine scenarios where rigid rule-following may be counterproductive. Designers should also consider how human-like behaviour and public acceptance intersect with safety and efficiency. Expert ID~11 highlighted that users are more likely to trust AVs that behave like human drivers, provided that safety is preserved. This has implications for user-centred AV design and for the development of regulatory frameworks that accommodate safe and socially acceptable deviations.

\vspace{0.75em}
\noindent\textbf{Limitations and Future Directions.}
This framework, while grounded in rich expert input, has limitations. First, expert reasoning may reflect regional or cultural biases. Second, generalisability across different AV scenarios—such as urban versus rural environments, or contexts with varying cultural norms—remains to be tested. Third, the development of future AV technologies, including Level4 and Level5 autonomy, will likely shift prioritisation patterns and override current constraints. Future work should empirically validate this framework across diverse contexts and conduct scenario-based testing with both users and developers. Longitudinal studies could explore how public expectations evolve over time and how AVs can adapt their decision-making while maintaining ethical and practical legitimacy.

\section{Conclusion} \label{sec:conclusion}

This study proposed a principled framework for AV behaviour in ethically ambiguous routine driving situations. Grounded in the tracking condition of Meaningful Human Control (MHC), the framework ensures that AV decisions align with human-centred reasons, such as values, intentions, and expectations. Through qualitative interviews with AV experts, we identified thirteen categories of reasons that influence manoeuvre planning, structured across normative, strategic, tactical, and operational levels of AV behaviour, and linked to the roles of relevant human agents.

The findings show that AV decisions often involve multiple overlapping reasons, with \textit{risk minimisation and safety assurance} consistently regarded as the primary concern. Other reasons—such as \textit{efficiency and trip optimisation}, \textit{comfort and user experience}, and \textit{user and public acceptance}—frequently accompany, but do not override, safety considerations. \textit{Regulatory compliance} was treated as conditional and often deprioritised when in tension with more context-sensitive goals. These prioritisation patterns support a framework that upholds safety while allowing for carefully constrained deviations from legal rules when justified by practical or ethical considerations.

By mapping expert reasons into a layered structure, the conceptual framework offers actionable design guidance that complements existing high-stakes ethical models and supports dynamic, real-world AV decision-making. Future research should validate this framework across diverse cultural contexts, automation levels, and driving scenarios.

\section{CRediT authorship and contribution statement}

\textbf{Lucas Elbert Suryana:} Conceptualization, Data curation, Formal analysis, Investigation, Methodology, Software, Validation, Visualization, Writing–original draft, Writing–review and editing.
\textbf{Simeon Calvert:} Conceptualization, Methodology, Supervision, Writing–review and editing.
\textbf{Arkady Zgonnikov:} Conceptualization, Methodology, Supervision, Writing–review and editing.
\textbf{Bart van Arem:} Conceptualization, Methodology, Supervision, Writing–review and editing.

\section{Acknowledgement}

We acknowledge the use of OpenAI ChatGPT to enhance the clarity and conciseness of, and to proofread, the manuscript; the original text was entirely authored by the researchers. This work was supported by the Indonesia Endowment Fund for Education (LPDP) under Grant No. 0006552/TRA/D/19/lpdp2021.


\appendix
\section{Questionnaire}
\label{sec:appendix:questionnaire}
\begin{longtable}{|l|p{0.7\linewidth}|}
\caption{Interview questions} \\
\hline
\textbf{Question number} & \textbf{Question} \\
\hline
\endfirsthead

\multicolumn{2}{c}%
{{\bfseries \tablename\ \thetable{} -- Continued from previous page}} \\
\hline
\textbf{Question number} & \textbf{Question} \\
\hline
\endhead

\hline \multicolumn{2}{r}{{Continued on next page}} \\
\endfoot

\hline
\endlastfoot
Q1 & Where do you work, what is your position and role, what are your activities with regard to automated vehicles (AVs)? \\
\hline
Q2 & What should automated vehicles (AVs) consider when planning a maneuver? Please give one example in much detail as possibl \\
\hline
Q3 & Which moral aspects do you believe AVs should consider when planning a maneuver? \\
\hline
Q4 & How might these aspects affect the manoeuvre plan? \\
\hline
\multicolumn{2}{|p{\dimexpr\linewidth-2\tabcolsep\relax}|}{
\textbf{Please watch the video below and read its description} \newline
\textbf{Video (see ~\ref{sec:appendix:video})}
\newline
\textbf{Video description}
\newline
A passenger uses an automated vehicle (AV) for a morning commute to the office. The passenger has an important meeting and must arrive on time. If the vehicle maintains the current speed, the passenger can reach the office on time in 20 minutes. The AV is on a road with solid double yellow lines, which prohibit vehicles from crossing in both directions due to safety reasons. During the trip, the AV approaches a cyclist traveling at half of the speed of the AV. There is no safe passing zone visible from the vehicle; however, the opposite lane is currently empty.  
} \\
\hline
Q5 & If the video continues, what do you believe all traffic participants will do? \\
\hline
Q6 & What are the reasons for the [traffic participants mentioned by the experts] performing the [actions the experts mentioned]? \\
\hline
Q7 & Besides the [traffic participants that are mentioned by the experts]'s, can you think any other factors that might influence the traffic participant decisions? \\
\hline
Q8 & What do you think the reasons are for the [other factors that are mentioned by the experts]? \\
\hline
Q9 & Can you think of any situations where the intentions of the [traffic participants / other factors the experts mentioned] might conflict? Please share any examples you can think of, and let me know when these conflicts may typically occur. \\
\hline
\multicolumn{2}{|p{\dimexpr\linewidth-2\tabcolsep\relax}|}{
\textbf{Recall the scene from the previous video.} \newline
There are three different people, each with their own intentions: \newline
\textbullet\ The automated vehicle (AV) passenger wants to pass the cyclist to get to the office on time. \newline
\textbullet\ The cyclist wants a safe distance from the AV for safety concerns. \newline
\textbullet\ The road policymaker wants both AV and cyclist to use their designated lanes, marked by solid yellow lines, for everyone's safety. \newline
Keep this in mind as you answer the rest of the questions.
} \\
\hline
Q10 & From your perspective, whose intentions should be given the most importance? Please answer this question by ranking the individuals below, with '1' indicating the highest rank. \\
     & \begin{tabular}{|l|c|c|c|}
     \hline
     & \textbf{1} & \textbf{2} & \textbf{3} \\
     \hline
     AV passenger & $\bigcirc$ & $\bigcirc$ & $\bigcirc$ \\
     Cyclist & $\bigcirc$ & $\bigcirc$ & $\bigcirc$ \\
     Road policymakers & $\bigcirc$ & $\bigcirc$ & $\bigcirc$ \\
     \hline
     \end{tabular} \\
\hline
Q11 & Could you please explain the reasons behind the rank you provided in your previous answer? \\
\hline
\multicolumn{2}{|p{\dimexpr\linewidth-2\tabcolsep\relax}|}{
\textbf{Watch the three video scenarios below!} \newline
These scenarios show three possible actions the AV might take if the previous video continues. The blue line ahead of the AV indicates the path it will follow. Imagine the AV's speed is the same in scenarios 2 and 3. The corresponding visual representations of these scenarios are available in ~\ref{sec:appendix:scenarios}.
\newline
\newline
\textbf{Scenario 1: AV stays behind the cyclist}
\newline
In this scenario, the AV only considers the cyclist's need for a safe distance and the road rules that require the AV to stay in its lane. But it doesn't consider the AV passenger's desire to get to the office on time.
\newline
\newline
\textbf{Scenario 2: AV overtakes the cyclist on its own lane}
\newline
In this scenario, the AV is solely concerned with the AV passenger's goal of getting to the office on time and the road rules that insist on it staying in its lane. But it doesn't consider the cyclist's wish to ride with a sense of safety.
\newline
\newline
\textbf{Scenario 3: AV overtakes the cyclist by using the opposite lane}
\newline
In this scenario, the AV is focused on the AV passenger's concern about getting to the office on time and the cyclist's concern about a safe distance. But it doesn't consider the road rules that require it to stay in its own lane.
}
\\
\hline
Q12 & Which of the above scenarios do you prefer? Please answer this question by ranking the scenarios, with '1' indicating the highest preference. \\
     & \begin{tabular}{|l|c|c|c|}
     \hline
     & \textbf{1} & \textbf{2} & \textbf{3} \\
     \hline
     Scenario 1: AV stays behind the cyclist & $\bigcirc$ & $\bigcirc$ & $\bigcirc$ \\
     Scenario 2: AV overtakes the cyclist on its own lane & $\bigcirc$ & $\bigcirc$ & $\bigcirc$ \\
     Scenario 3: AV overtakes the cyclist by using the opposite lane & $\bigcirc$ & $\bigcirc$ & $\bigcirc$ \\
     \hline
     \end{tabular} \\
\hline
Q13 & Could you please explain the reasons behind the rank you provided in your previous answer? \\
\hline
\multicolumn{2}{|p{\dimexpr\linewidth-2\tabcolsep\relax}|}{
\textbf{Take a look at the video below!} \newline
\newline
\textbf{Scenario 4: AV overtakes the cyclist by crossing some part of the opposite lane}
\newline
In this scenario, the AV only considers the cyclist's need for a safe distance and the road rules that require the AV to stay in its lane. But it doesn't consider the AV passenger's desire to get to the office on time. See the time instance visuals in ~\ref{sec:appendix:scenario4}.
\newline\newline
You will now assess how much you believe the AV considers the intentions of three different stakeholders across 7 moments in this scenario. The same response table below will be used to answer Questions 14, 16, and 18. See the corresponding time instance visuals in ~\ref{sec:appendix:scenario4}.
\newline
\begin{tabular}{|l|c|c|c|c|c|c|c|c|c|c|c|}
     \hline
     & \textbf{0} & \textbf{10} & \textbf{20} & \textbf{30} & \textbf{40} & \textbf{50} & \textbf{60} & \textbf{70} & \textbf{80} & \textbf{90} & \textbf{100}\\
     \hline
     Time instance 1 & $\bigcirc$& $\bigcirc$ & $\bigcirc$ & $\bigcirc$ & $\bigcirc$ & $\bigcirc$ & $\bigcirc$ & $\bigcirc$ & $\bigcirc$ & $\bigcirc$ & $\bigcirc$ \\
     Time instance 2 & $\bigcirc$& $\bigcirc$ & $\bigcirc$ & $\bigcirc$ & $\bigcirc$ & $\bigcirc$ & $\bigcirc$ & $\bigcirc$ & $\bigcirc$ & $\bigcirc$ & $\bigcirc$ \\
     Time instance 3 & $\bigcirc$& $\bigcirc$ & $\bigcirc$ & $\bigcirc$ & $\bigcirc$ & $\bigcirc$ & $\bigcirc$ & $\bigcirc$ & $\bigcirc$ & $\bigcirc$ & $\bigcirc$ \\
     Time instance 4 & $\bigcirc$& $\bigcirc$ & $\bigcirc$ & $\bigcirc$ & $\bigcirc$ & $\bigcirc$ & $\bigcirc$ & $\bigcirc$ & $\bigcirc$ & $\bigcirc$ & $\bigcirc$ \\
     Time instance 5 & $\bigcirc$& $\bigcirc$ & $\bigcirc$ & $\bigcirc$ & $\bigcirc$ & $\bigcirc$ & $\bigcirc$ & $\bigcirc$ & $\bigcirc$ & $\bigcirc$ & $\bigcirc$ \\
     Time instance 6 & $\bigcirc$& $\bigcirc$ & $\bigcirc$ & $\bigcirc$ & $\bigcirc$ & $\bigcirc$ & $\bigcirc$ & $\bigcirc$ & $\bigcirc$ & $\bigcirc$ & $\bigcirc$ \\
     Time instance 7 & $\bigcirc$& $\bigcirc$ & $\bigcirc$ & $\bigcirc$ & $\bigcirc$ & $\bigcirc$ & $\bigcirc$ & $\bigcirc$ & $\bigcirc$ & $\bigcirc$ & $\bigcirc$ \\
     
     \hline
     \end{tabular}}
\\
\hline
Q14 & Imagine you are the AV passenger in scenario 4. Please assess, for each time instance, how much you believe the AV considers your intention to arrive at the office on time. (0 = Not consider at all; 100 = Fully consider) \\
\hline
Q15 & Please clarify why the scores you provided for each time instance are either constant or change at each time instance \\
\hline
Q16 & Imagine you are the cyclist that is passed by the AV in scenario 4. Please assess, for each time instance, how much you believe the AV considers your intention to bike with a sense of safety. (0 = Not consider at all; 100 = Fully consider) \\
\hline
Q17 & Please clarify why the scores you provided for each time instance are either constant or change at each time instance \\
\hline
Q18 & Imagine you are a road policymaker, and you see the AV briefly occupying small part of the opposite lane when overtaking the cyclist in scenario 4. Please assess, for each time instance, how much you believe the AV considers your intention putting a solid double yellow lines on the road for safety reasons. (0 = Not consider at all; 100 = Fully consider) \\
\hline
Q19 & Please clarify why the scores you provided for each time instance are either constant or change at each time instance \\
\hline
Q20 & How can an AV understands the intention of people on the road and other stakeholders in a real traffic situation? \\
\hline
Q21 & From your point of view, what approach should AVs use to manage possible conflicts between the intentions of people on the road and other stakeholders in real traffic situations? \\
\hline
\end{longtable}

\section{Clip of Video Scenario}
\label{sec:appendix:video}

\begin{figure}[H]
  \centering
  \includegraphics[width=0.5\linewidth]{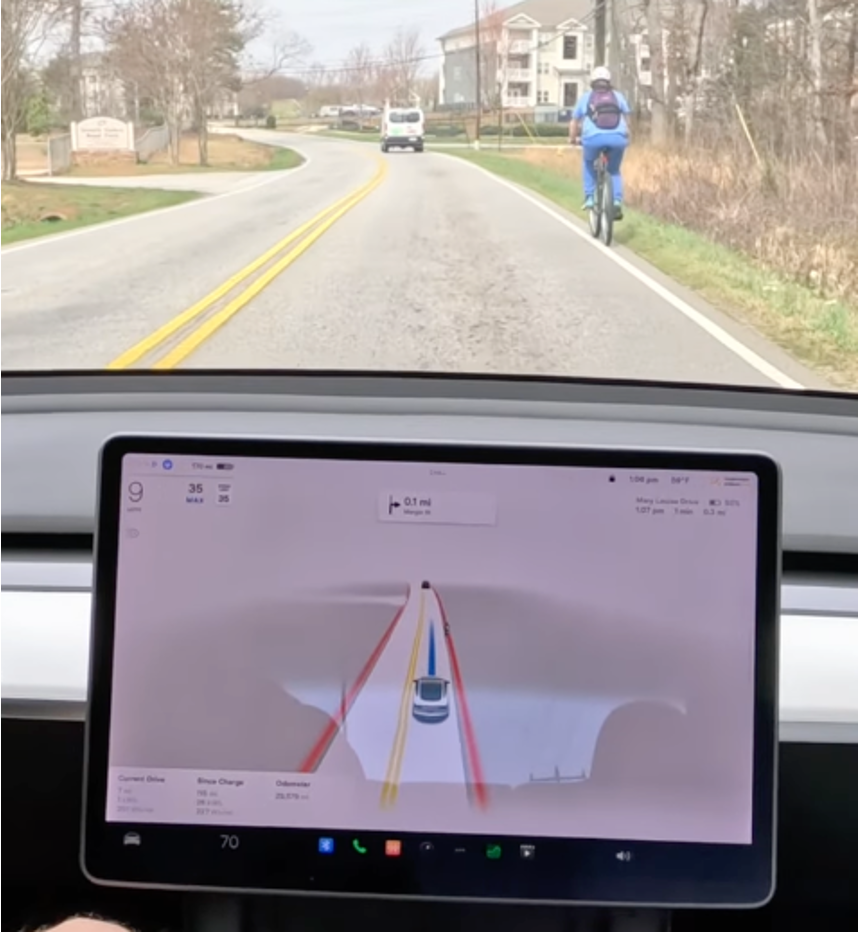}
  \caption{Source: adapted and edited by the author based on original footage \citep{TeslaShort2023}.}
  \label{fig:video_scenario}
\end{figure}

\section{Scenario Clips for Questions 12–13}
\label{sec:appendix:scenarios}

\begin{figure}[H]
  \centering
  \includegraphics[width=0.6\linewidth]{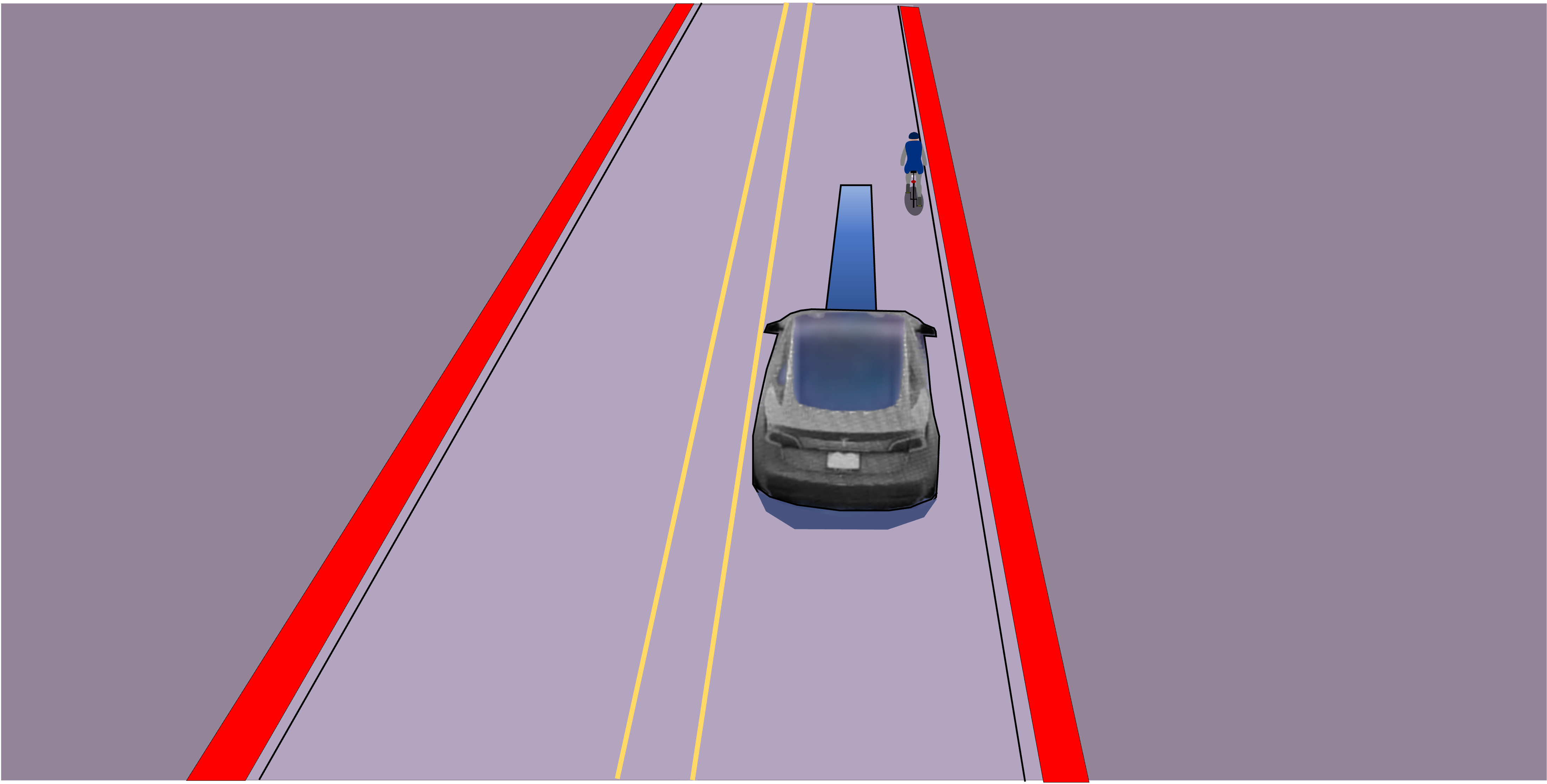}
  \caption{Scenario 1: AV stays behind the cyclist.}
  \label{fig:scenario1}
\end{figure}

\begin{figure}[H]
  \centering
  \includegraphics[width=0.6\linewidth]{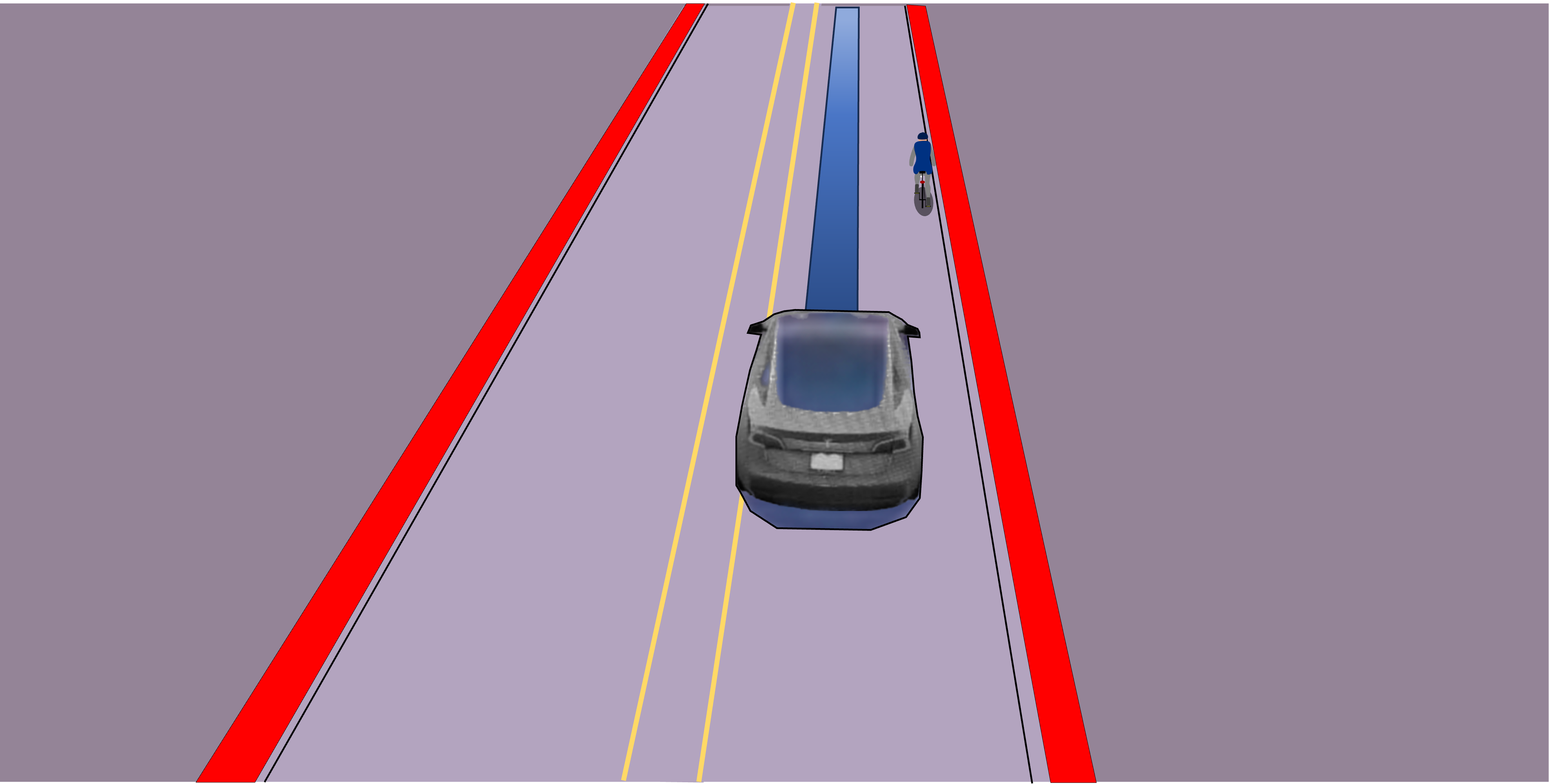}
  \caption{Scenario 2: AV overtakes the cyclist in its own lane.}
  \label{fig:scenario2}
\end{figure}

\begin{figure}[H]
  \centering
  \includegraphics[width=0.6\linewidth]{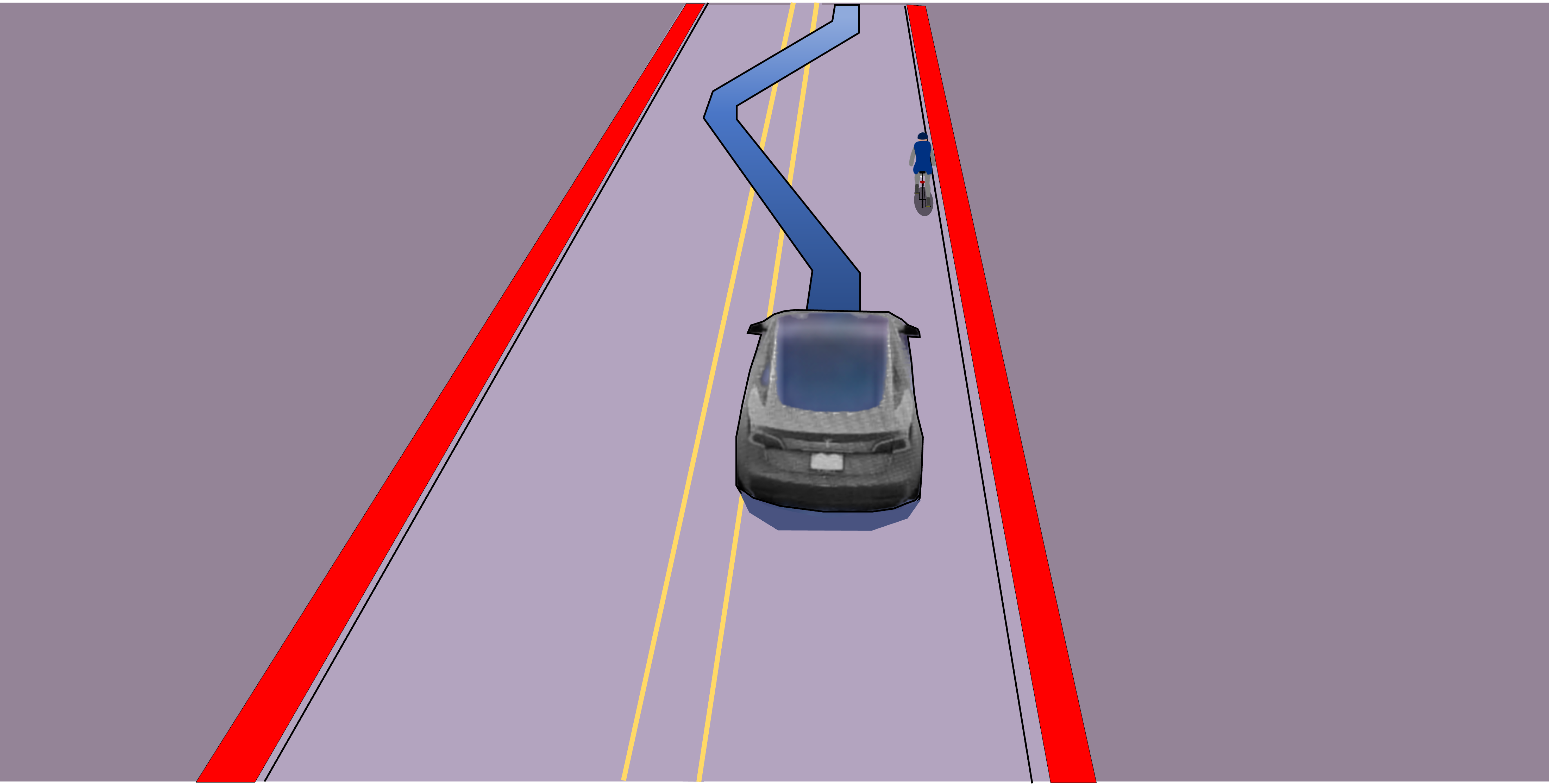}
  \caption{Scenario 3: AV overtakes the cyclist using the opposite lane.}
  \label{fig:scenario3}
\end{figure}

\section{Scenario 4 Time Instances}
\label{sec:appendix:scenario4}

\begin{figure}[H]
  \centering
  \includegraphics[width=0.6\linewidth]{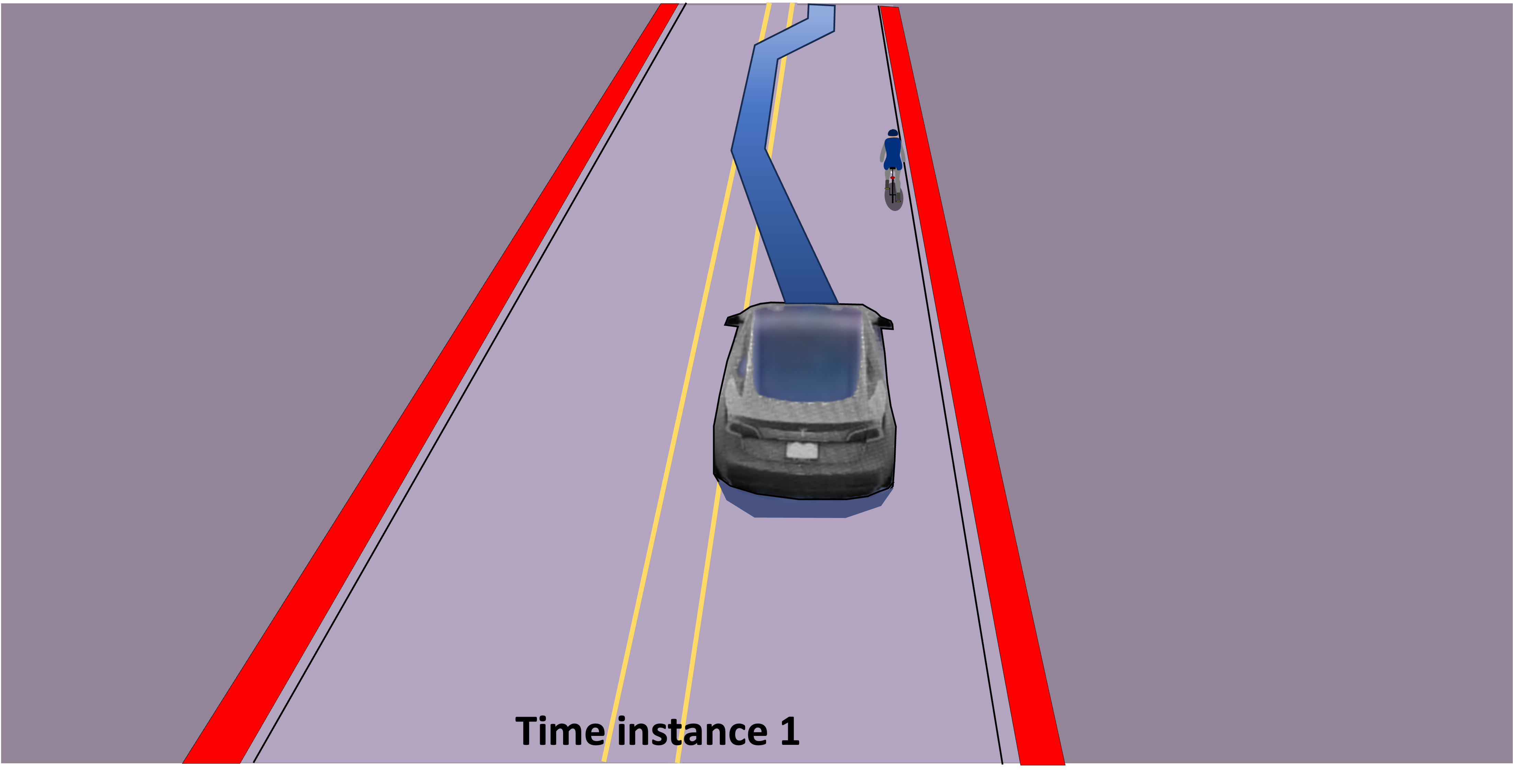}
  \caption{Time instance 1}
\end{figure}

\begin{figure}[H]
  \centering
  \includegraphics[width=0.6\linewidth]{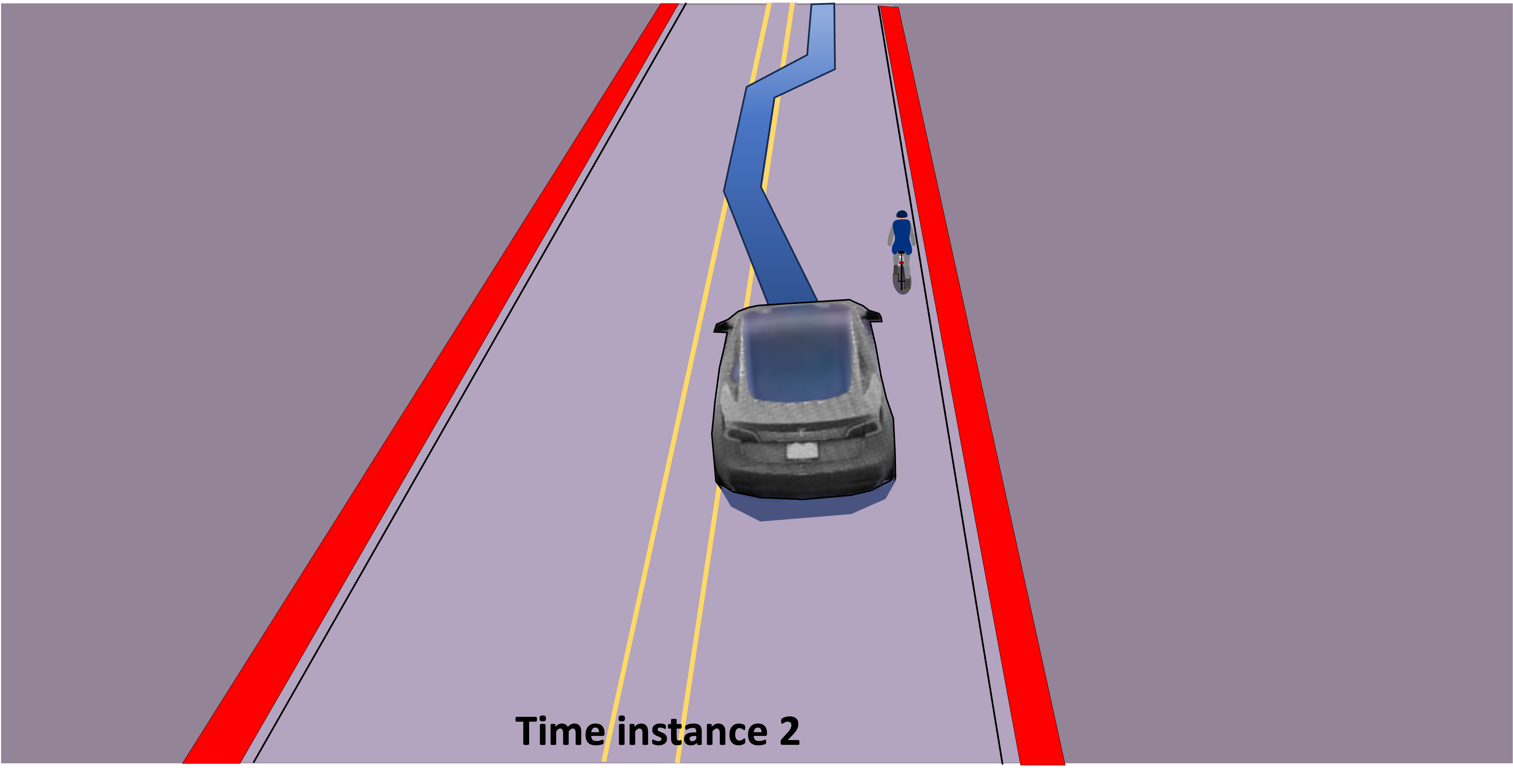}
  \caption{Time instance 2}
\end{figure}

\begin{figure}[H]
  \centering
  \includegraphics[width=0.6\linewidth]{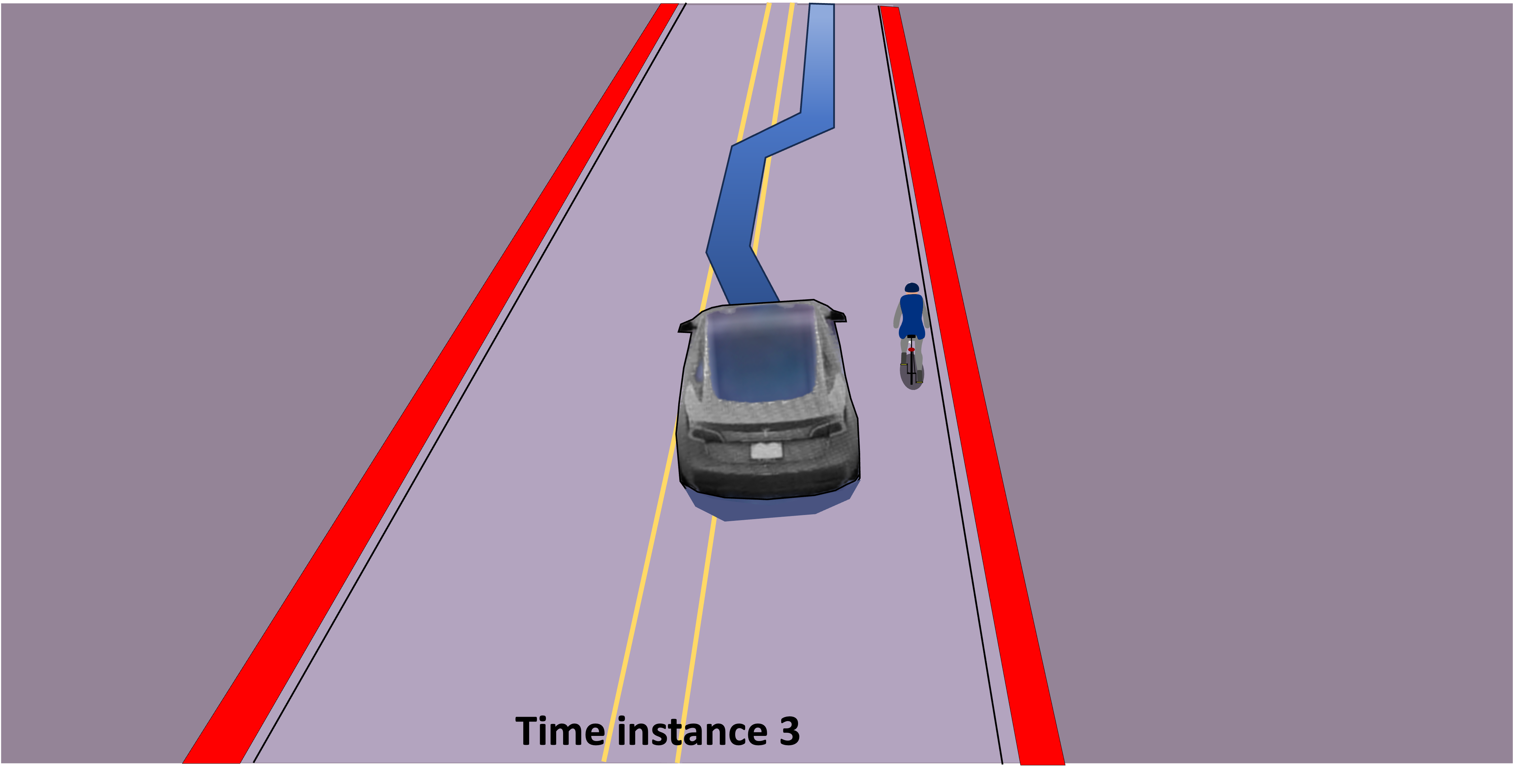}
  \caption{Time instance 3}
\end{figure}

\begin{figure}[H]
  \centering
  \includegraphics[width=0.6\linewidth]{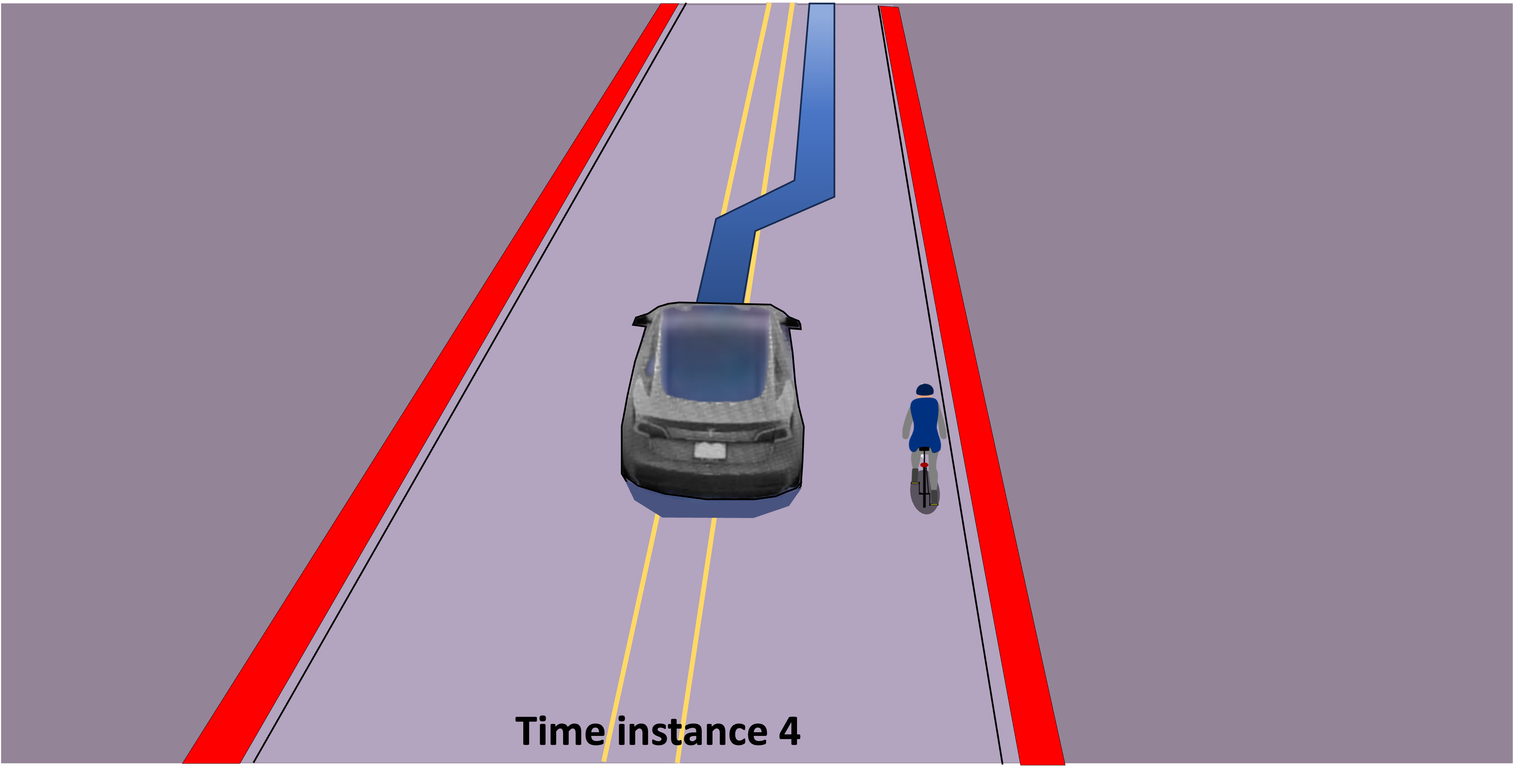}
  \caption{Time instance 4}
\end{figure}

\begin{figure}[H]
  \centering
  \includegraphics[width=0.6\linewidth]{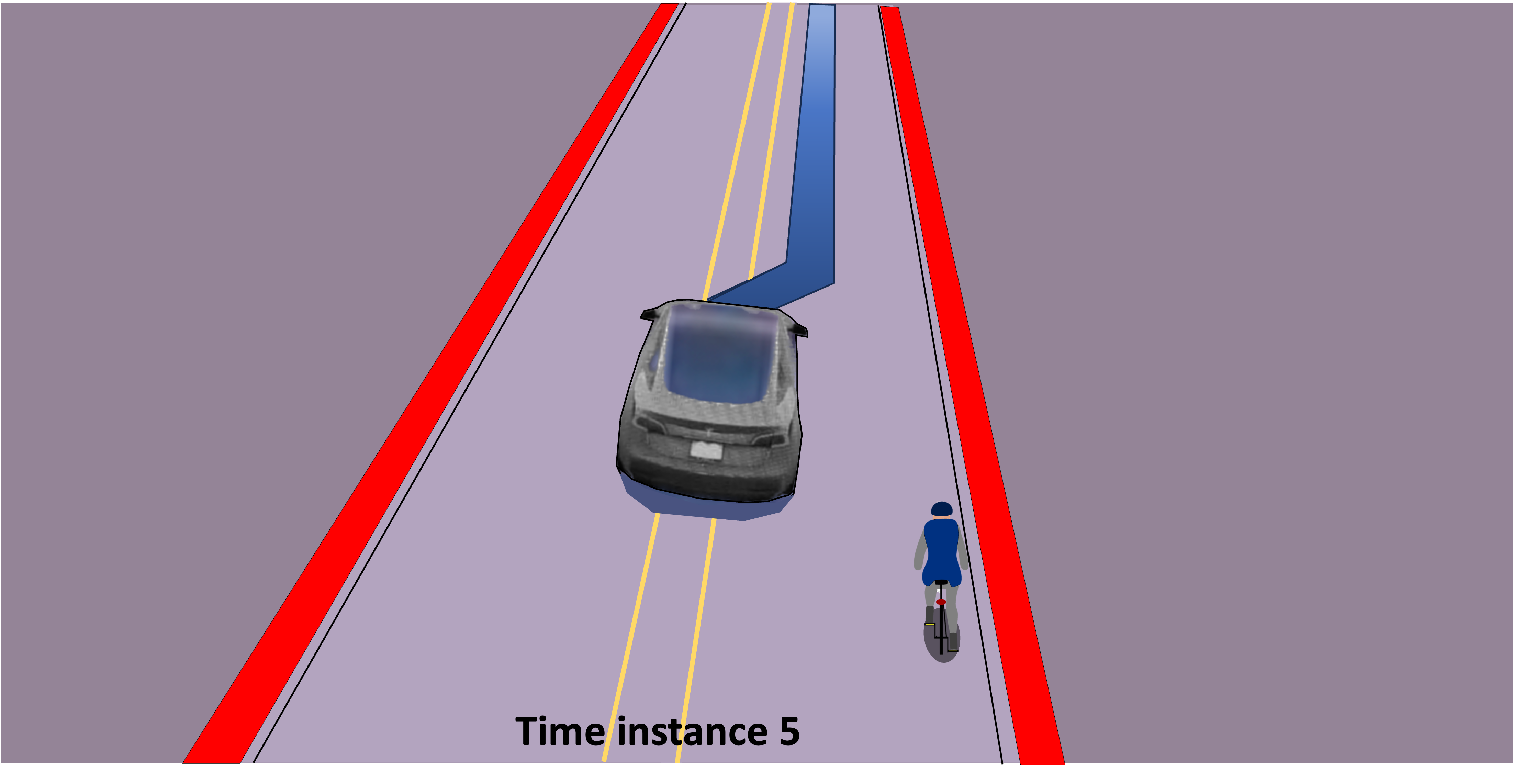}
  \caption{Time instance 5}
\end{figure}

\begin{figure}[H]
  \centering
  \includegraphics[width=0.6\linewidth]{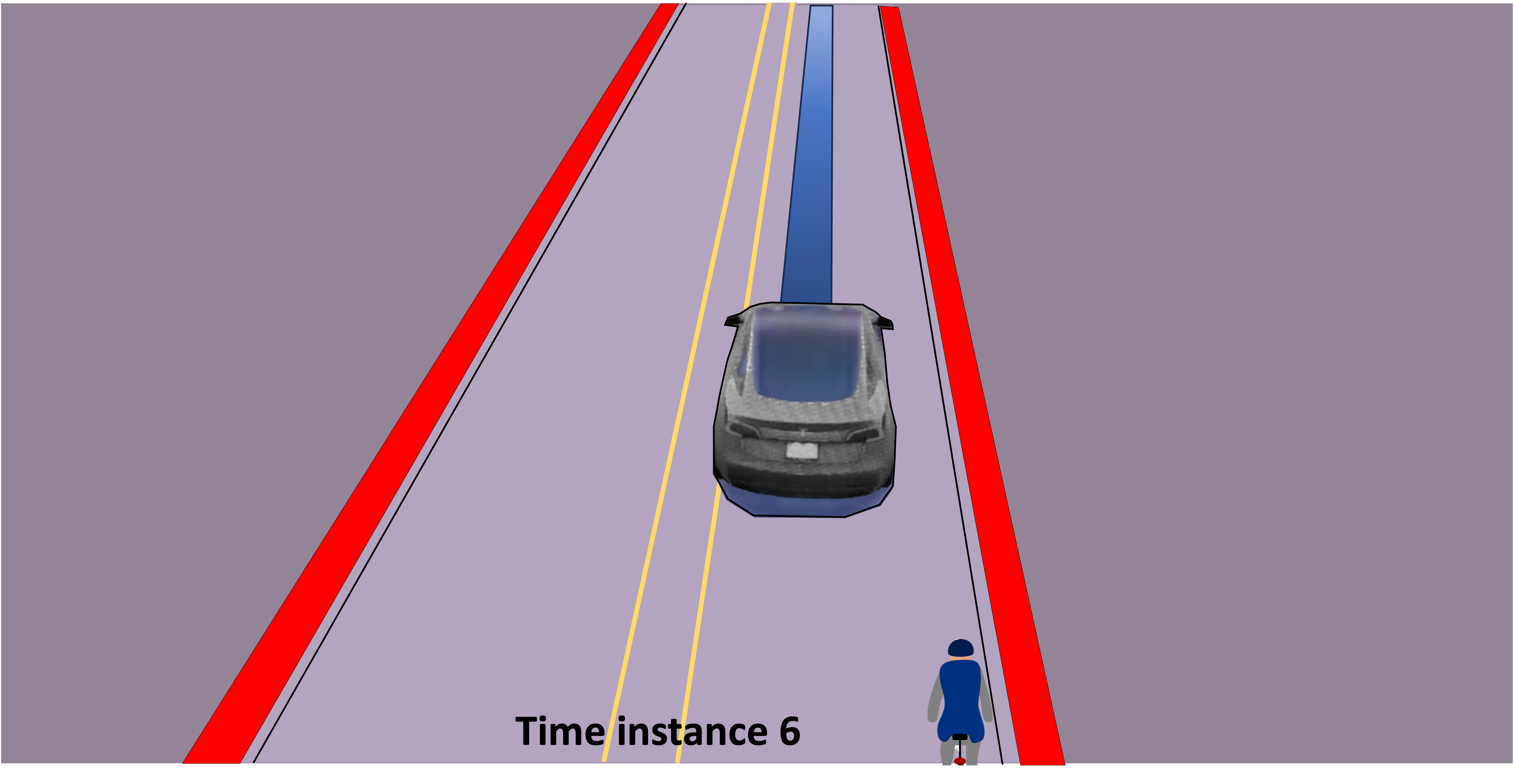}
  \caption{Time instance 6}
\end{figure}

\begin{figure}[H]
  \centering
  \includegraphics[width=0.6\linewidth]{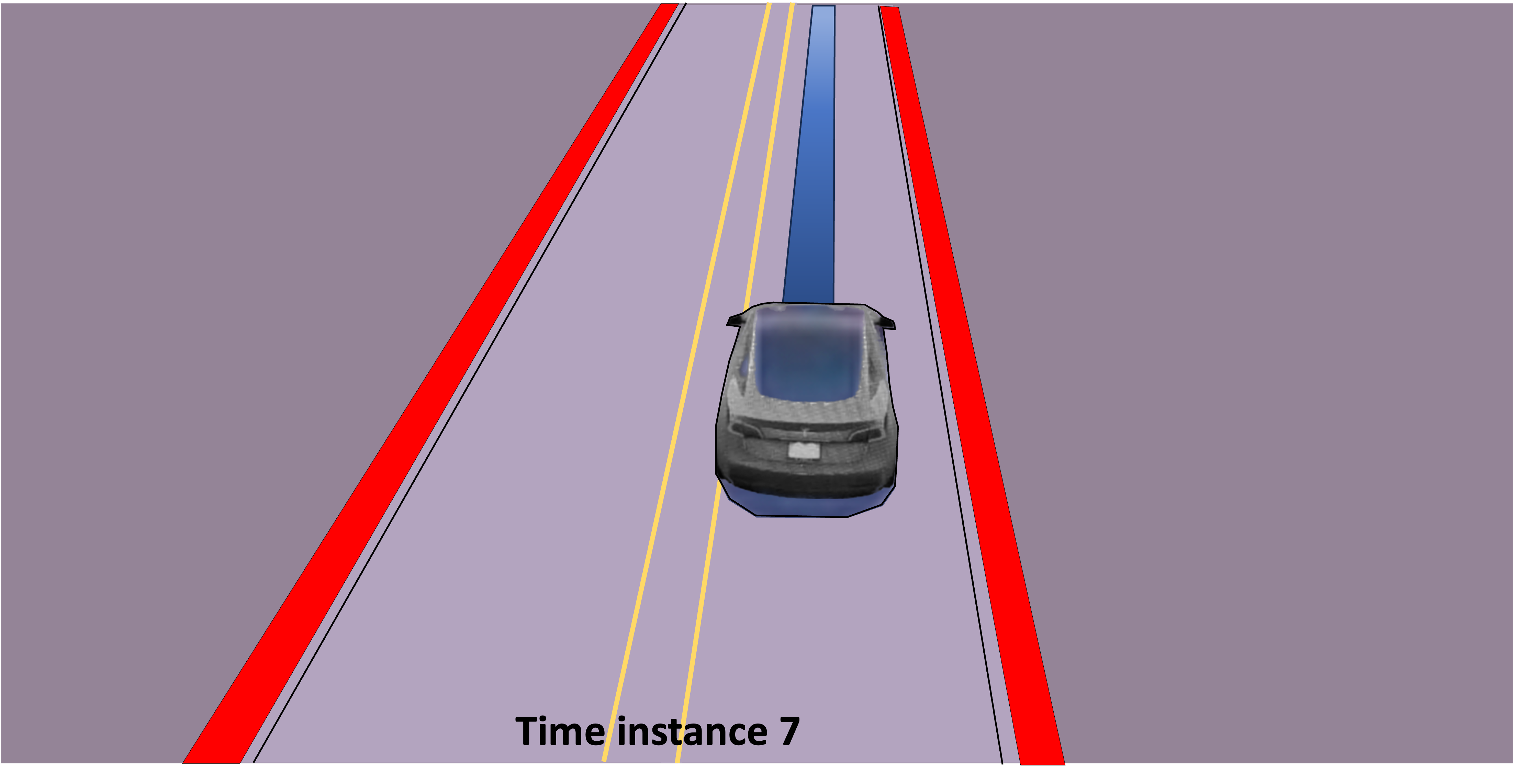}
  \caption{Time instance 7}
\end{figure}







\end{document}